\documentclass[aps,prd,showpacs,superscriptaddress,preprintnumbers]{revtex4}
\usepackage{graphicx,float,wrapfig,subfigure}
\usepackage{amsfonts,amsmath,amssymb,amstext}
\usepackage{latexsym}
 \usepackage{bm}
\usepackage{color}
\usepackage[normalem]{ulem}

\newcommand{\be}{\begin{equation}}
\newcommand{\ee}{\end{equation}}
\newcommand{\ba}{\begin{eqnarray}}
\newcommand{\ea}{\end{eqnarray}}

\newcommand{\sign}{\,\mbox{sign}}

\definecolor{red}{rgb}{0.7,0,0}
\definecolor{green}{rgb}{0,0.5,0}

\begin{document}

\title{Ellipticity of photon emission from strongly magnetized hot QCD plasma}
\date{\today}

\author{Xinyang Wang}
\email{wangxy@ujs.edu.cn}
\affiliation{Department of Physics, Jiangsu University, Zhenjiang 212013 P.R. China}

\author{Igor A. Shovkovy}
\email{igor.shovkovy@asu.edu}
\affiliation{College of Integrative Sciences and Arts, Arizona State University, Mesa, Arizona 85212, USA}
\affiliation{Department of Physics, Arizona State University, Tempe, Arizona 85287, USA}

\author{Lang Yu}
\affiliation{College of Physics, Jilin University, Changchun 130012 P.R. China}

\author{Mei Huang}
\affiliation{School of Nuclear Science and Technology, University of Chinese Academy of Sciences, Beijing 100049, P.R.China}

\begin{abstract}
By making use of an explicit representation for the imaginary part of the photon polarization tensor in terms of  
transitions between the Landau levels of light quarks, we study the angular dependence of direct photon emission 
from a strongly magnetized quark-gluon plasma. Because of the magnetic field, the leading order photon rate 
comes from the three processes of the zeroth order in the coupling constant $\alpha_s$: (i) the quark splitting 
($q\rightarrow q+\gamma $), (ii) the antiquark splitting ($\bar{q} \rightarrow \bar{q}+\gamma $), and 
(iii) the quark-antiquark annihilation ($q + \bar{q}\rightarrow \gamma$). In a wide range of moderately high 
temperatures, $T\gtrsim m_\pi$, and strong magnetic fields, $|eB|\gtrsim m_\pi^2$, the direct photon production 
is dominated by the two splitting processes. We show that the Landau-level quantization of quark states plays 
an important role in the energy and angular dependence of the photon emission. Among other things, it leads 
to a nontrivial momentum dependence of the photon ellipticity coefficient $v_2$, which takes negative values 
at small transverse momenta and positive values at large transverse momenta. The crossover between the two 
regimes occurs around $k_T\simeq \sqrt{|eB|}$. In application to heavy-ion collisions, this suggests that a large 
value of $v_2$ for the direct photons could be explained in part by the magnetic field in the quark-gluon plasma.
\end{abstract}
\maketitle

\section{Introduction}
\label{sec:introduction}

For the last half a century, there has been a growing interest in the problem of strongly interacting QCD matter under extreme conditions. One of the extreme regimes, which is characterized by a high energy density, is the quark-gluon plasma (QGP) produced in heavy-ion collision experiments at the Relativistic Heavy Ion Collider (RHIC) in Brookhaven and the Large Hadron Collider (LHC) at CERN. While the deconfined QCD matter produced in relativistic heavy-ion collisions is initially far from equilibrium, it is likely to approach a quasiequilibrium state on a relatively short timescale. There is also a growing consensus that the resulting strongly interacting QGP behaves almost like a perfect hydrodynamic fluid. Because of the high initial pressure, the plasma expands rapidly and its temperature decreases. Thus, a detailed study of the corresponding evolution could be used to shed light on a large part of the QCD phase diagram. Such knowledge is not only of interest in heavy-ion physics but may also provide an insight into the physics of the early Universe.

High temperature is not the only extreme feature of the QGP produced in heavy-ion collisions. In the case of noncentral collisions, in particular, the resulting QCD matter is also characterized by a super-strong magnetic field~\cite{Skokov:2009qp,Deng:2012pc} and very large vorticity~\cite{Jiang:2016woz,Deng:2016gyh}. The exploration of  such unusual conditions is of fundamental interest because both magnetic field and vorticity could trigger a range of interesting anomalous phenomena. The chiral magnetic effect (CME)~\cite{Fukushima:2008xe,Kharzeev:2007tn,Kharzeev:2007jp} and the chiral vortical effect (CVE)~\cite{Kharzeev:2007tn,Son:2009tf,Kharzeev:2010gr} are perhaps the most popular among them. Theoretically, such effects can modify dramatically the collective behavior of relativistic matter. In heavy-ion collision experiments, the effects can be revealed by detailed studies of multiparticle correlators of charged particles and the spin polarization of neutral particles. The anomalous phenomena in question are of fundamental interest since they promise the possibility of extracting anomalous quantum effects from bulk properties of matter. As the recent progress in the field suggests, the same anomalous physics can be relevant also for applications in astrophysics, cosmology, and even for a class of topological semimetals, e.g., see Ref.~\cite{Miransky:2015ava}.

The hydrodynamics features of the QGP in heavy-ion collisions are supported by the measurements of the anisotropic flow coefficients. In essence, the latter are the eccentricity (Fourier) coefficients of multiparticle correlators averaged over many events. Theoretically, an anisotropic flow of plasma is seeded by the initial spatial asymmetry of the overlap region of nuclei colliding with a nonzero impact parameter. Because of the unavoidable event-by-event fluctuations in collisions, as well as large statistical fluctuations due to the small system size, all flow coefficients are expected to be nonzero. At midrapidity, however, the second harmonics $v_2$ (describing the average ellipticity of flow) is expected to be particularly important and informative. Indeed, the latter should be dominated by the initial pressure anisotropy stemming from an almond-shape overlap region of the colliding nuclei. 

One of the curious observations in the heavy-ion experiments at RHIC and LHC is a strong azimuthal asymmetry of the photon production in a wide  range of rapidities. The first measurement of the elliptic flow of direct photons in Au-Au collisions was reported by the PHENIX collaboration at RHIC nearly a decade ago~\cite{Adare:2011zr}. Later the same collaboration published more precise measurements with an extension to lower values of the transverse momentum~\cite{Adare:2015lcd}. The photon elliptic flow of a similar magnitude was also reported independently by the ALICE collaboration at LHC~\cite{Acharya:2018bdy}. The most surprising fact was the magnitude of the photon flow which is comparable to the flow of hadrons. To explain the experimental data, a barrage of theoretical studies was triggered~\cite{Chatterjee:2005de,Schenke:2006yp,Chatterjee:2008tp,vanHees:2011vb,Linnyk:2013wma,Gale:2014dfa,Muller:2013ila,vanHees:2014ida,Monnai:2014kqa,Dion:2011pp,Liu:2012ax,Vujanovic:2014xva,McLerran:2014hza,McLerran:2015mda,Gelis:2004ep,Hidaka:2015ima,Linnyk:2015rco,Vovchenko:2016ijt,Koide:2016kpe,Turbide:2005bz,Tuchin:2012mf,Basar:2012bp}. Nevertheless, it is fair to say that the current understanding of underlying physics is still far from being complete. 

It is reasonable to expect that a strong magnetic field, produced in noncentral heavy-ion collisions, can affect the photon emission \cite{Yee:2013qma,Tuchin:2014pka,Zakharov:2016mmc}. Since the magnetic field is likely to be present during an extended period of the evolution of the fireball \cite{Skokov:2009qp,Deng:2012pc,Tuchin:2015oka,Guo:2019mgh}, all known sources of photon emission could be affected. Here we will concentrate primarily on the direct photon emission from the quark-gluon plasma. The latter is the dominant mechanism during the early stages of quark-gluon plasma when the magnetic field is particularly strong. As we show, quantum transitions between the Landau levels of quarks lead to photon emission with unique properties. 

We argue that, in a strongly magnetized plasma, the photon rate is dominated by the following three single-photon processes: (i) the quark splitting ($q\rightarrow q+\gamma $), (ii) the antiquark splitting ($\bar{q} \rightarrow \bar{q}+\gamma $), and (iii) the quark-antiquark annihilation ($q + \bar{q}\rightarrow \gamma$). This is in contrast to the vanishing magnetic field case when these three processes are forbidden by the energy-momentum conservation. In fact, the leading order result at $B=0$ is given by the gluon-mediated $2\to 2$ processes $q+g\rightarrow q+\gamma $, $\bar{q} +g \rightarrow \bar{q}+\gamma $, and $q + \bar{q}\rightarrow g+ \gamma$, where $g$ represents a gluon \cite{Kapusta:1991qp,Baier:1991em,Aurenche:1998nw,Steffen:2001pv,Arnold:2001ba,Arnold:2001ms,Ghiglieri:2013gia}. 
This explains why the corresponding leading-order photon rate is linear in the strong coupling constant $\alpha_s$. In the presence of a strong magnetic field, in contrast, the $1\to 2$ splitting  and $2\to 1$ annihilation processes are allowed by the energy-momentum conservation. As a result, the photon rate is nonzero already at the leading zeroth order in $\alpha_s$. The corresponding rate is calculated explicitly in this paper.

From the general considerations, it is clear that the presence of a strong magnetic field could trigger a strong emission of direct photons. Moreover, the rate is expected to have a rather nontrivial dependence on the magnitude and direction of the photon momentum. As we show in this study, the Landau level quantization plays an important role in the emission of photons with small transverse momenta (i.e, $k_T \lesssim \sqrt{|eB|}$). Also, the emission rate tends to be the highest in the directions along the line of the magnetic field (i.e., perpendicularly to the reaction plane). While the quantization is less important at large transverse momenta (i.e, $k_T \gtrsim \sqrt{|eB|}$), the emission still has a strong dependence on the direction relative to the magnetic field. In fact, similarly to the classical synchrotron radiation, the preferred direction of the photon emission at large $k_T$ is perpendicular to the magnetic field (i.e., in the reaction plane) \cite{Tuchin:2014pka,Zakharov:2016mmc}.  

This paper is organized as follows. In Sec.~\ref{sec:anisotropic-flow}, we give a general overview of the elliptic flow in heavy-ion experiments. In Sec.~\ref{sec:Polarization}, the explicit expression for the imaginary part of the one-loop photon polarization tensor is presented. Numerical results for the direct photon production and the ellipticity of emission in a magnetized plasma are presented in Sec.~\ref{sec:Numerical-results}. The summary of the main results and conclusions are given in Sec.~\ref{sec:summary}. Some technical details and calculations are provided in several appendices at the end of the paper.

\section{Anisotropic flow}
\label{sec:anisotropic-flow}

In noncentral heavy-ion collisions, the anisotropic flow coefficients ($v_n$) are defined by the following Fourier decomposition of the azimuthal particle distributions \cite{Voloshin:1994mz,Poskanzer:1998yz}:
\begin{equation}
E \frac{{\mathrm d}^3 N}{{\mathrm d}^3 \mathbf{p}} = 
\frac{1}{2\pi} \frac {{\mathrm d}^2N}{p_{T}{\mathrm d}p_{T}{\mathrm d} y} 
  \left(1 + 2\sum_{n=1}^{\infty}  v_n \cos[n(\phi-\Psi_{\rm RP})]\right), 
\label{vnor}
\end{equation}
where $E$ is the particle energy, $\mathbf{p}$ is the momentum, $p_{T}$ is 
the transverse momentum, $\phi$ is the azimuthal angle, $y$ is  the rapidity,
and $\Psi_{\rm RP}$ is the reaction plane angle. By definition, 
\begin{equation}
v_n(p_T,y) = \langle \cos[n(\phi-\Psi_{\rm RP})] \rangle,
\label{fouriercoeff}
\end{equation}
where the angular brackets denote the average over all particles (or all events, or both) in a given bin of the transverse momentum ($p_T$) and rapidity ($y$). Note that the first two coefficients in the Fourier decomposition (\ref{vnor}), i.e., $v_1$ and $v_2$, characterize the directed  and elliptic flow, respectively.

\begin{figure}[t]
\centering
\includegraphics[width=0.45\textwidth]{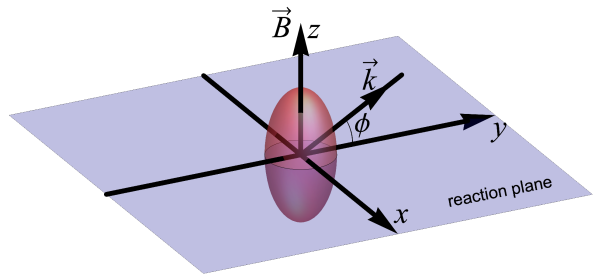}
\caption{A schematic illustration of the reaction plane and the coordinate system used.}
\label{illustration}
\end{figure}

As shown schematically in Fig.~\ref{illustration}, the direction of the magnetic field in a noncentral collision is (approximately) perpendicular to the reaction plane. In the study below, we will assume that the corresponding direction is the $z$ axis of the coordinate system used. Also, by assumption, $x$-$y$ is the reaction plane and the $x$ axis points along the beam direction. The azimuthal angle $\phi$ measures the angle between the photon momentum $\mathbf{k}$ and the reaction plane. The photon four-momentum is given by $k^\mu=(k^0,\mathbf{k})$. Note that the transverse components of the photon momentum are given by
\begin{equation}
\label{kT-phi-parametrization}
k_y = k_T \cos(\phi),\quad k_z =k_T \sin(\phi).
\end{equation}
where $k_T=\sqrt{k_y^2+k_z^2}$ is the magnitude of the transverse momentum. Here we set $k_x = 0$ which corresponds to the case of midrapidity ($y=0$). 

By making use of the general representation in Eq.~(\ref{vnor}), the differential distribution of photons is given by 
\begin{equation}
k^0 \frac{d^3R}{dk_x dk_y dk_z}  = \frac{d^3 R}{k_{T} d k_T d\phi dy} = \frac{1}{2\pi}  \frac{d^2 R}{k_{T} d k_T  dy} \left[1+\sum_{n=1}^{\infty}2v_n(k_T, y) \cos(n\phi)\right],
\label{diff-rate-1}
\end{equation}
where we used $dk_x= k_0 dy$, which follows from the definitions $k_0=\sqrt{k_T^2+k_x^2}$ and $y=\frac{1}{2}\ln\frac{k_0+k_x}{k_0-k_x}$.

By making use of quantum field theory, the thermal photon production rate can be expressed in terms of the imaginary part of the retarded polarization tensor as follows \cite{Kapusta:2006pm}:
\begin{equation}
k^0\frac{d^3R}{dk_x dk_y dk_z}=-\frac{1}{(2\pi)^3}\frac{\mbox{Im}\left[\Pi^{\mu}_{\mu}(k)\right]}{\exp\left(\frac{k_0}{T}\right)-1},
\label{diff-rate-2}
\end{equation}
As usual, this assumes that the mean free path of photons is larger that the system size so that the photon leave the plasma region without reabsorption.

\begin{figure}[t]
\centering
\includegraphics[width=0.3\textwidth]{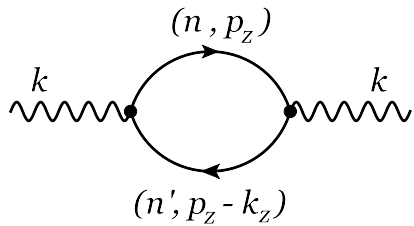}
\caption{The one-loop Feynman diagram for the photon polarization tensor in a magnetic field.}
\label{polarization}
\end{figure}

Note that, to the leading one-loop order, the photon polarization tensor is given by the Feynman diagram in Fig.~\ref{polarization}, where the internal solid lines represent quark propagators in a background magnetic field. Here, by considering a sufficiently strong magnetic field, we  assume that the higher-loop diagrams with gluon-mediated interactions will produce subleading corrections. It should be emphasized, however, that the interplay between the thermal and magnetic effects could be quite nontrivial in general. 

It is easy to see from the definition in Eq.~(\ref{diff-rate-1}) that the anisotropy coefficients $v_n$ can be evaluated from the differential distribution of photons as follows:
\begin{equation}
v_n = \frac{1}{\mathcal{R}_0}  \int_0^{2\pi} \frac{d^3 R}{k_{T} d k_T d\phi dy}  \cos(n \phi) d \phi ,
\end{equation}
where the normalization factor is given by the photon production rate integrated over the angular coordinate $\phi$, i.e.,
\begin{equation}
\mathcal{R}_0 = \frac{d^2 R}{k_{T} d k_T  dy} = \int_0^{2\pi} \frac{d^3 R}{k_{T} d k_T  dy d\phi} d \phi .
\end{equation}
Similarly, by making use of the quantum field theoretical expression in Eq.~(\ref{diff-rate-2}), one can extract the anisotropy coefficients as follows:
\begin{equation}
v_n(k_T) = - \frac{1}{(2\pi)^3\mathcal{R}}  \int_0^{2\pi}\frac{ \mbox{Im}\left[\Pi^{\mu}_{\mu}(k)\right]}{\exp\left(\frac{k_0}{T}\right)-1} \cos(n\phi)d\phi,
\label{v2}
\end{equation}
where the corresponding normalization factor is defined by
\begin{equation}
\mathcal{R} = - \frac{1}{(2\pi)^3 } \int_0^{2\pi}\frac{ \mbox{Im}\left[\Pi^{\mu}_{\mu}(k)\right]}{\exp\left(\frac{k_0}{T}\right)-1} d\phi .
\label{Integrated-rate}
\end{equation}
Below we will use the definition in Eq.~(\ref{v2}) to determine the ellipticity ($v_2$) of the direct photon production in a hot magnetized quark-gluon plasma.

\section{Polarization function}
\label{sec:Polarization}

The photon polarization tensor in the presence of a background magnetic field was studied by a number of authors. At zero temperature, the most comprehensive studies were reported in Refs.~\cite{Hattori:2012je,Hattori:2012ny}. Some studies of the  polarization tensor have been done also at nonzero temperature. In particular, the results in the lowest Landau level approximation were obtained in Refs.~\cite{Miransky:2002rp,Bandyopadhyay:2016fyd}, and in the weak field limit in Refs.~\cite{Das:2019nzv,Ghosh:2019kmf}. Several interesting results have been also obtained by using the Ritus method and the real time formalism in Refs.~\cite{Sadooghi:2016jyf} and~\cite{Ghosh:2018xhh}. In this paper, to study the ellipticity of the direct photon emission, we will use an explicit expression for the imaginary part of the polarization tensor obtained in Ref.~\cite{Wang-Shovkovy:2020}. The corresponding result has a relatively simple form and a clear interpretation in terms of quantum transitions between quantized Landau levels of light quarks. 

By omitting most of technical details of the derivation in Ref.~\cite{Wang-Shovkovy:2020}, it is instructive to discuss the underlying assumptions and highlight the key steps that lead to the final expression for the imaginary part of the polarization tensor $\mbox{Im}[\Pi^{\mu}_{\mu}(k)]$. To leading order in coupling, the photon polarization tensor is given by the flavor sum of the one-loop Feynman diagrams shown in Fig.~\ref{polarization}, where the internal lines represent the quark propagators in a background magnetic field. 
We will assume that the contributions of the lightest up and down quarks dominate the photon polarization function in the quark-gluon plasma at moderately high temperatures. For simplicity, we will also assume that the masses of the light quarks are the same, i.e., $m_u = m_d =m$. While the strange quark is not taken into account, its inclusion is straightforward if needed. In either case, the role of strange quark is not critical for the purposes of the current study, which examines the qualitative features of the direct photon emission from a strongly magnetized plasma. 

We will assume that the magnetic field $\mathbf{B}$ points in the $+z$ direction and the vector potential is given by the Landau gauge, i.e., $\mathbf{A}=(-B y,0,0)$. In such a background field, the quark propagator takes the following form~\cite{Miransky:2015ava}:
\begin{equation}
G_f(t-t^\prime;\mathbf{r},\mathbf{r}^\prime) = e^{i\Phi^f (\mathbf{r}_\perp,\mathbf{r}_\perp^\prime)}\bar{G}_f(t-t^\prime;\mathbf{r}-\mathbf{r}^\prime), 
\label{quark-prop}
\end{equation}
where $\mathbf{r}_\perp=(x,y)$ is the transverse coordinate (in the reaction plane), $f=u,d$ is the flavor index, and $\Phi^{f}(\mathbf{r}_\perp,\mathbf{r}_\perp^\prime)=-e_f B(x-x^{\prime})(y+y^{\prime})/2$ is the well-known Schwinger phase. According to our conventions here, $e_f = q_f e$ is the quark charge, where $q_u = 2/3$, $q_d = -1/3$, and $e$ is the absolute value of the electron charge. 

It is convenient to rewrite the translation invariant part of the propagator $\bar{G}_f$ in Eq.~(\ref{quark-prop}) by using the following  mixed coordinate-momentum space representation~\cite{Miransky:2015ava}:
\begin{equation}
\bar{G}_f(t;\mathbf{r}) = \int \frac{d\omega dp_z }{(2\pi)^2} e^{-i\omega t +i p_z z}
\bar{G}_f(\omega; p_z ;\mathbf{r}_\perp) ,
\end{equation}
where 
\begin{equation}
\bar{G}_f(\omega, p_z ;\mathbf{r}_\perp) = i\frac{e^{-\mathbf{r}_\perp^2/(4l_f^2)}}{2\pi l_f^2}
\sum_{n=0}^{\infty}
\frac{\tilde{D}^f_{n}(\omega,p_z ;\mathbf{r}_\perp)}{\omega^2-E_{n,p_z,f}^2},
\label{GDn-alt}
\end{equation}
and $E_{n,p_z,f}= \sqrt{m^2+p_z^2 + 2 n |e_f B|}$ is the quark energy in the $n$th Landau level. In the last expression, we also
used the following shorthand notation for the numerator of the $n$th Landau-level contribution~\cite{Miransky:2015ava}:
\begin{equation}
\tilde{D}_{n}^f(\omega,p_z ;\mathbf{r}_\perp) = \left[\omega\gamma^0 -p^{3}\gamma^3 + m \right]
\left[\mathcal{P}^f_{+}L_n\left(\frac{\mathbf{r}_{\perp}^2}{2l^{2}_f}\right)
+\mathcal{P}^f_{-}L_{n-1}\left(\frac{\mathbf{r}_{\perp}^2}{2l^{2}_f}\right)\right]
-\frac{i}{l_f^2}(\mathbf{r}_{\perp}\cdot\bm{\gamma}_{\perp}) 
 L_{n-1}^1\left(\frac{\mathbf{r}_{\perp}^2}{2l_f^{2}}\right),
 \label{Dn-quark}
\end{equation}
where $L_{n}^\alpha(z)$ are the generalized Laguerre polynomials \cite{Gradshtein},
$\mathcal{P}^f_{\pm}\equiv \frac12 \left(1\pm i s^f_\perp \gamma^1\gamma^2\right)$ 
are spin projectors, and $l_f=\sqrt{1/|e_f B|}$ is the flavor-specific  magnetic length. By definition, 
$s_\perp^f=\sign (e_f B)$ and $L_{-1}^\alpha(z) \equiv 0$. 

The finite-temperature expression for the polarization function is given by
\begin{equation}
\Pi^{\mu\nu}(i\Omega_m;\mathbf{k}) =  4\pi N_c \sum_{f = u, d} \alpha_f T \sum_{k=-\infty}^{\infty} \int \frac{dp_z}{2\pi} 
\int d^2 \mathbf{r}_\perp e^{-i \mathbf{r}_\perp\cdot \mathbf{k}_\perp} 
\mbox{tr} \left[ \gamma^\mu \bar{G}_{f}(i\omega_k, p_z ;\mathbf{r}_\perp)  
\gamma^\nu \bar{G}_{f}(i\omega_k-i\Omega_m, p_z-k_z; -\mathbf{r}_\perp)\right],
\label{Pi_Omega_k-alt}
\end{equation}
where $\alpha_f = q_f^2 \alpha$,  $\alpha= e^2/(4\pi)$ is the fine structure constant, $N_c=3$ is the number of colors, and the trace on the right-hand side runs over the Dirac indices. Note that the photon and quark Matsubara frequencies are given by 
$\Omega_m = 2\pi m T$ and $\omega_k = \pi (2k+1) T$, respectively. 

After summing over the Matsubara frequencies, the retarded polarization tensor is obtained from the thermal Green's function by replacing $i \Omega_m$ with $\Omega+ i \epsilon$. (Here we use the notation $k_0 =\Omega $ for the photon.) The imaginary part of the corresponding (Lorentz-contracted) polarization function reads~\cite{Wang-Shovkovy:2020}:
\begin{eqnarray}
\mbox{Im} \left[\Pi_{R,\mu}^{\mu}(\Omega;\mathbf{k}) \right] &=&
\sum_{f=u, d} \frac{N_c\alpha_f}{2l_f^4} \sum_{n,n^\prime=0}^{\infty}\int \frac{dp_z}{2\pi} 
\sum_{\lambda,\eta=\pm 1} 
\frac{n_F(E_{n,p_z,f})-n_F(\lambda E_{n^{\prime},p_z-k_z,f}) }{2 \eta\lambda  E_{n,p_z,f}E_{n^{\prime},p_z-k_z,f}}
\sum_{i=1}^{4}\mathcal{F}_{i}^{f}
\nonumber\\
&\times&
\delta\left(E_{n,p_z,f}-\lambda E_{n^{\prime},p_z-k_z,f}+\eta \Omega\right).
\label{Im-Pol-fun}
\end{eqnarray}
where the explicit expressions for functions $\mathcal{F}_{i}^{f}$ are given in Appendix~\ref{all-functions}. 
From the physics viewpoint, it is important to note that the $\delta$ function has a nonvanishing support only 
when the energy conservation equation $E_{n,p_z,f}-\lambda E_{n^{\prime},p_z-k_z,f}+ \eta\Omega=0$ is 
satisfied. 

It is instructive to remember that, unlike the real part of the polarization function, the imaginary part should 
have no ultraviolet divergencies. This is confirmed by a careful analysis of the explicit expression in 
Eq.~(\ref{Im-Pi-final}), where the sum over Landau levels is convergent. Depending on the model parameters, 
however, the inclusion of a large number of terms could be required for a reliable evaluation of the imaginary 
part. 

\begin{figure}[t]
\centering
  \subfigure[]{\includegraphics[width=0.25\textwidth]{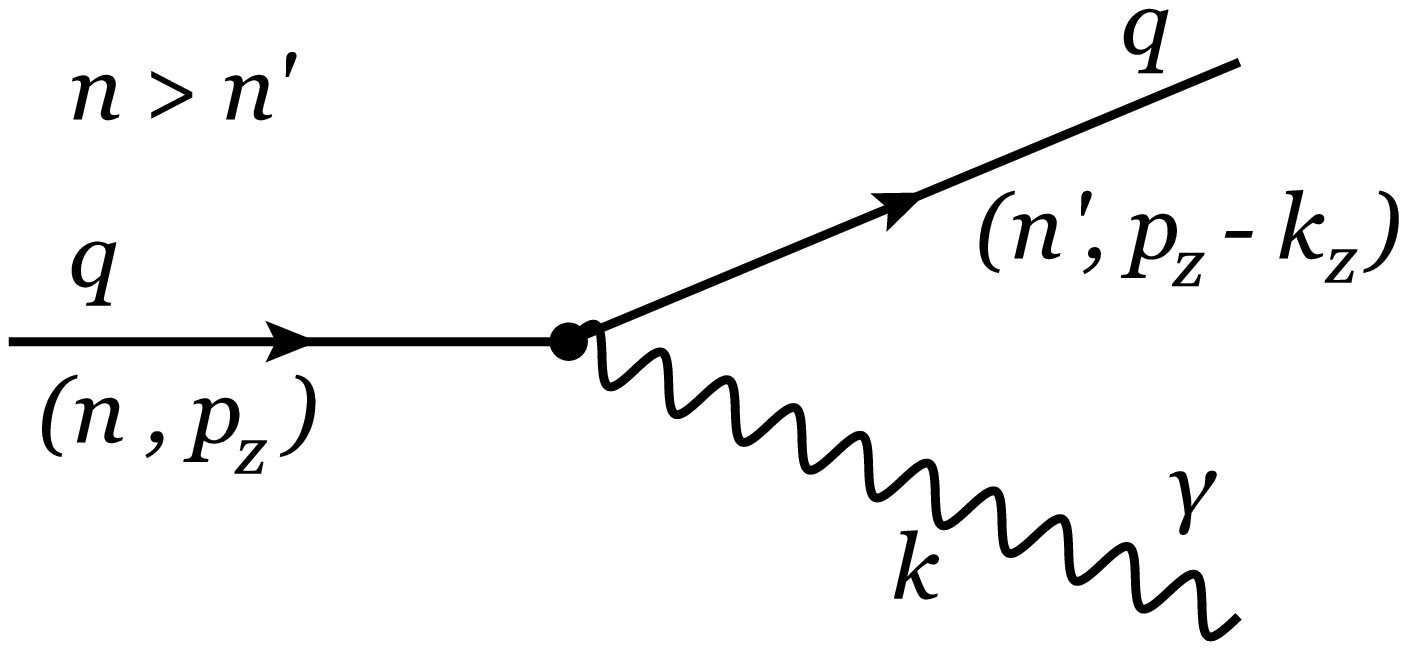}}
  \hspace{0.1\textwidth}
  \subfigure[]{\includegraphics[width=0.25\textwidth]{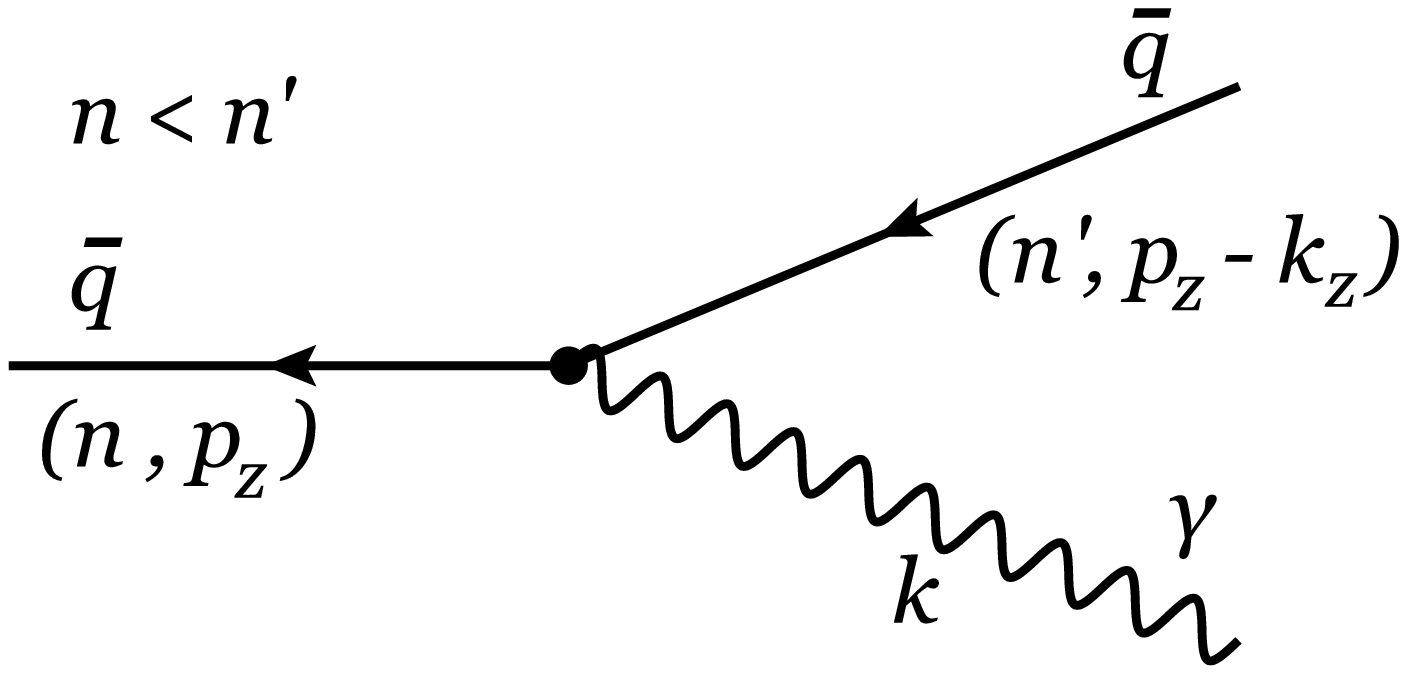}}
  \hspace{0.1\textwidth}
  \subfigure[]{\includegraphics[width=0.25\textwidth]{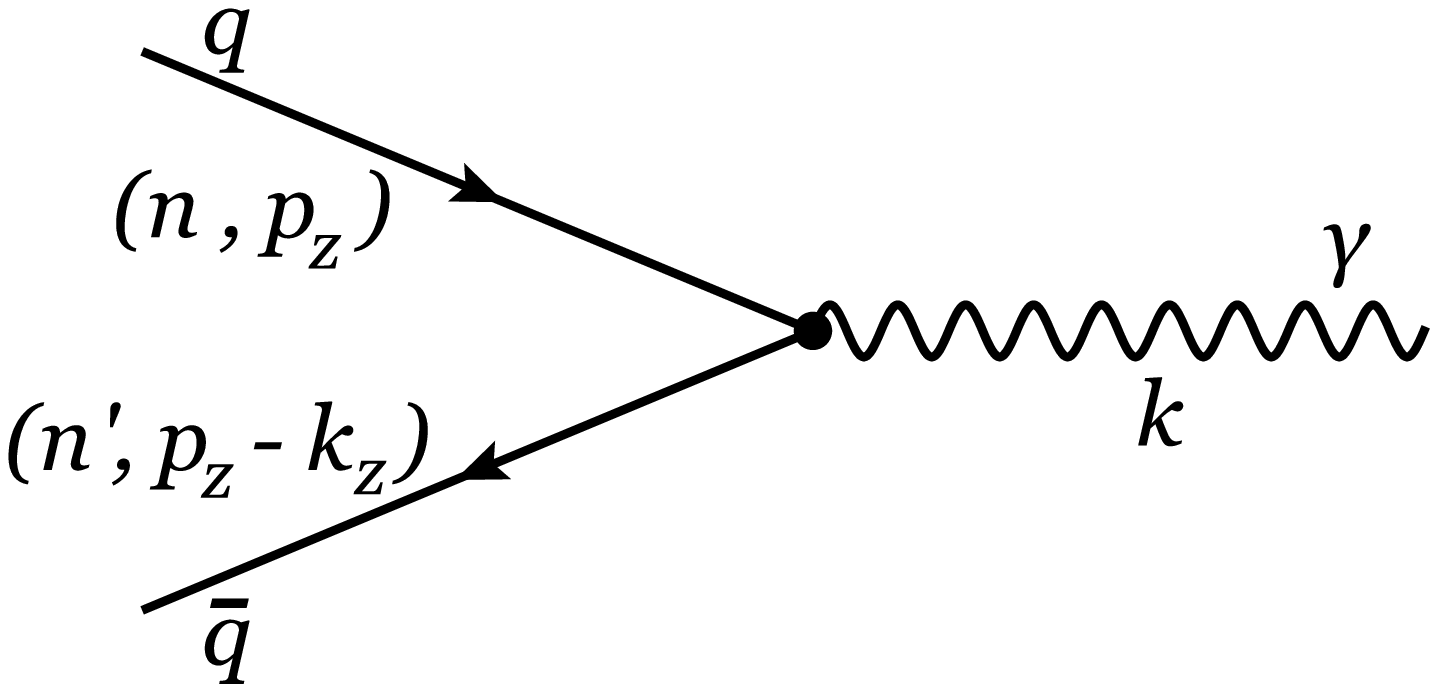}}\\
  \subfigure[]{\includegraphics[width=0.25\textwidth]{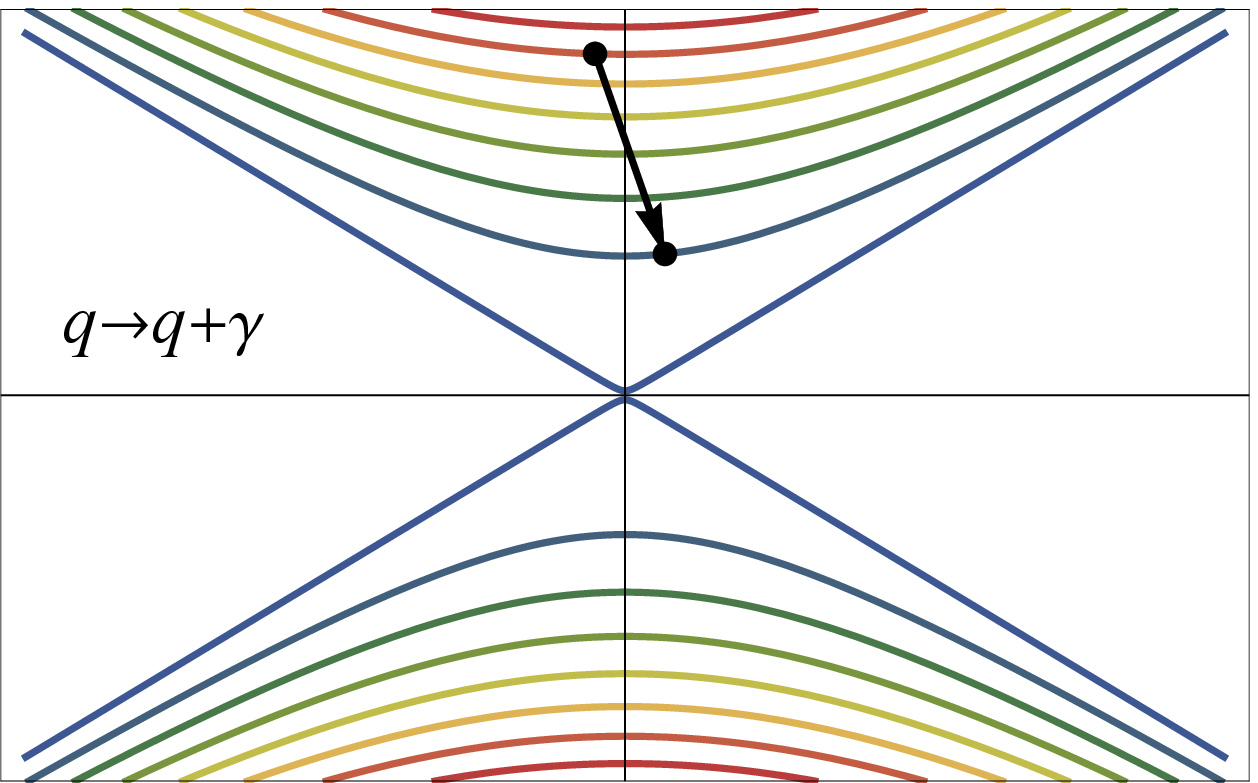}}
  \hspace{0.1\textwidth}
  \subfigure[]{\includegraphics[width=0.25\textwidth]{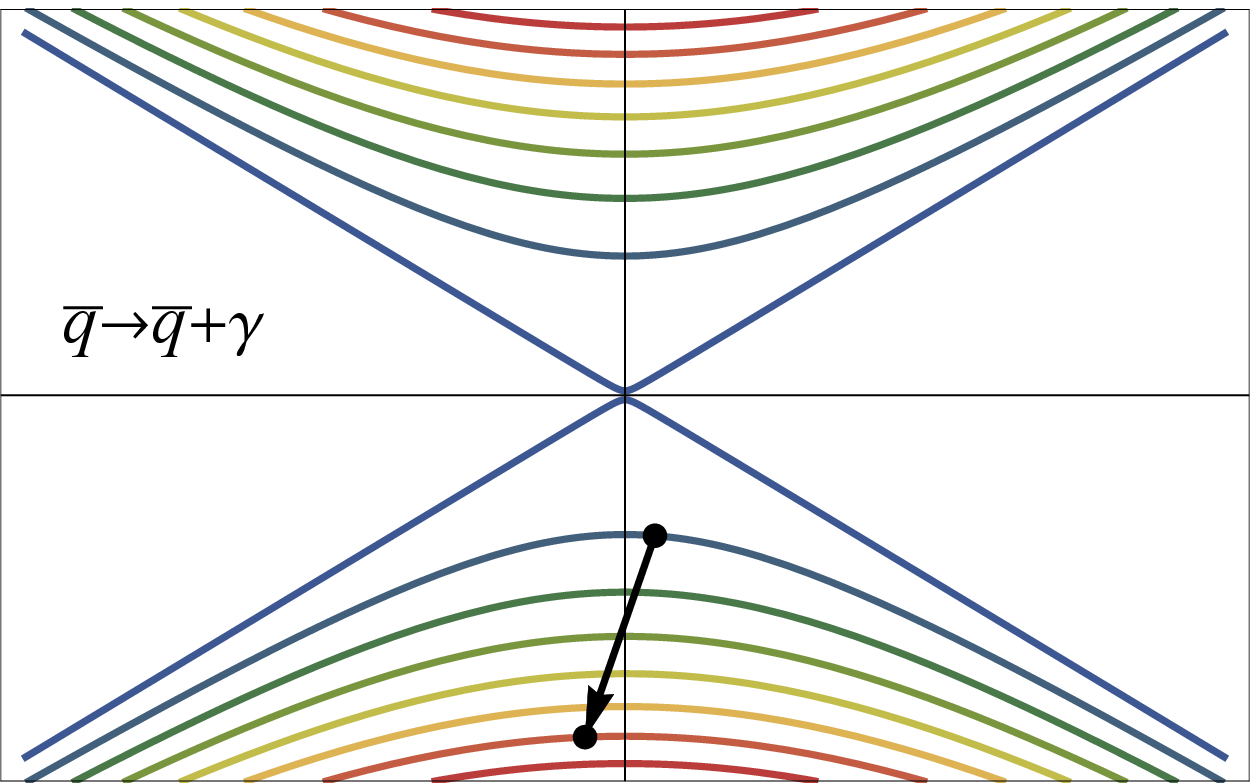}}
  \hspace{0.1\textwidth}
  \subfigure[]{\includegraphics[width=0.25\textwidth]{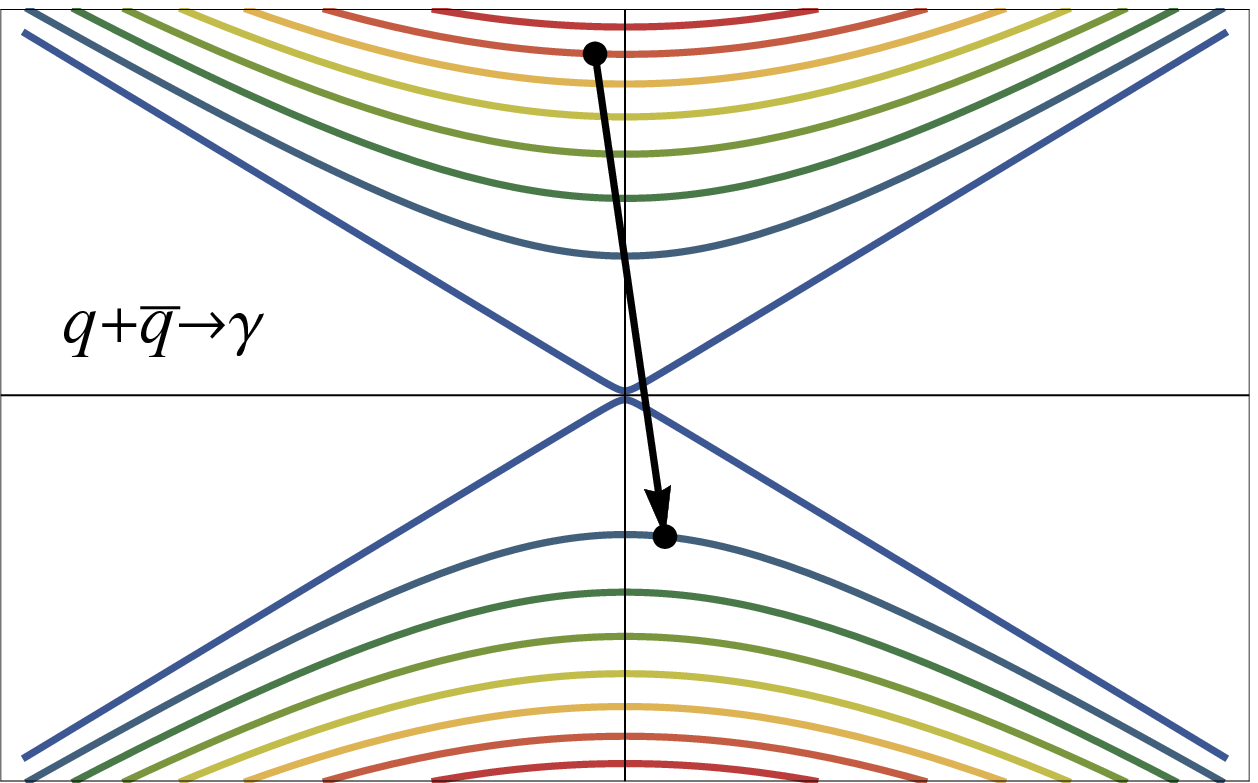}}  
\caption{Three types of processes involving fermion states with the Landau-level indices $n$ and $n^{\prime}$:
(a) $q\to q+\gamma$, (b) $\bar{q}\to \bar{q}+\gamma$, (c) $q+ \bar{q}\to \gamma$. The corresponding Landau 
level transitions are shown schematically in panels (d), (e) and (f), respectively.}
\label{LLprocesses}
\end{figure}

Without loss of generality, let us assume that $\Omega$ is positive. Taking into account that the 
imaginary part should be an odd function of $\Omega$, this is not a strong limitations. (By making use of the 
time reversal transformation, it is also natural to associate negative values of $\Omega$ with the photon 
absorption processes. Such an interpretation is also supported by the analysis of the energy 
conservation relation.) Then, depending on the choice of signs of $\lambda$ and 
$\eta$, the energy conservation equation $E_{n,p_z,f}-\lambda E_{n^{\prime},p_z-k_z,f}+ \eta\Omega=0$ 
represents one of the three possible physical processes involving quark and/or antiquark states with the 
Landau-level indices $n$ and $n^{\prime}$. Two of the processes, which are realized when $\lambda =+1$, 
are the quark and antiquark splitting processes, i.e., $q\to q +\gamma$ ($\eta =-1$) and 
$\bar{q}\to \bar{q} +\gamma$  ($\eta =+1$),  respectively. The third possibility is the annihilation process 
$q+ \bar{q}\to \gamma$, which is realized when $\lambda =-1$ and $\eta = - 1$. It is easy to verify that 
there are no physical processes that correspond to $\lambda = -1$ and $\eta = +1$ when  $\Omega$ is positive. 

The existence of real solutions for $p_z$ to the energy conservation equation implies that the corresponding 
process is allowed in principle. In the case of three types of processes mentioned above, the necessary 
conditions for the existence of real solutions are given as follows:
\begin{eqnarray}
q\to q +\gamma~(\lambda =+1,~ \eta =-1): &\quad&
\sqrt{\Omega^2-k_z^2} \leq k_{-}^{f}
\mbox{~and~} n>n^{\prime} ,
\label{qqg-cond}
\\
\bar{q}\to \bar{q} +\gamma~ (\lambda =+1,~ \eta =+1): &\quad&
\sqrt{\Omega^2-k_z^2} \leq k_{-}^{f}
\mbox{~and~} n<n^{\prime} ,
\label{bqbqg-cond}
\\
q+ \bar{q}\to \gamma~(\lambda =-1,~ \eta =-1): &\quad&
\sqrt{\Omega^2-k_z^2} \geq k_{+}^{f} ,
\label{qbqg-cond}
\end{eqnarray}
where we utilized the shorthand notation
\begin{equation}
k_{\pm}^{f} = \left|\sqrt{m^2+2n|e_fB|} \pm \sqrt{m^2+2n^{\prime}|e_fB|}\right| .
\end{equation}
When the appropriate condition is satisfied for a given process, see Eqs.~(\ref{qqg-cond}) -- (\ref{qbqg-cond}) 
the real solutions for $p_z$ are given by the following explicit expressions:
\begin{eqnarray}
p_{z,f}^{(\pm)}&=& \frac{k_z}{2}\left[1+ \frac{2(n-n^{\prime})|e_fB|}{\Omega^2-k_z^2} 
\pm \frac{\Omega}{|k_z|}  \sqrt{ \left[1-\frac{(k_{-}^f)^2}{\Omega^2-k_z^2} \right]
\left[ 1-\frac{(k_{+}^f)^2}{\Omega^2-k_z^2}\right]} \right].
\label{pz-solution}
\end{eqnarray}
On these solutions, the quark energies take the following explicit forms:
\begin{eqnarray}
 \left. E_{n,p_z,f}\right|_{p_z = p_{z,f}^{(\pm)} } &=& - \frac{\eta \Omega}{2} \left[
1+\frac{2(n-n^{\prime})|e_fB|}{\Omega^2-k_z^2}\pm \frac{|k_z|}{\Omega}\sqrt{ \left(1-\frac{(k_{-}^{f})^2}{\Omega^2-k_z^2} \right)\left( 1-\frac{(k_{+}^{f})^2}{\Omega^2-k_z^2}\right)} 
\right], \label{E1-solution}\\
 \left. E_{n^{\prime},p_z-k_z,f} \right|_{p_z = p_{z,f}^{(\pm)} } &=& \frac{\lambda \eta \Omega}{2} \left[
1-\frac{2(n-n^{\prime})|e_fB|}{\Omega^2-k_z^2}\mp \frac{|k_z|}{\Omega}\sqrt{ \left(1-\frac{(k_{-}^{f})^2}{\Omega^2-k_z^2} \right)\left( 1-\frac{(k_{+}^{f})^2}{\Omega^2-k_z^2}\right)} 
\right],
\label{E2-solution}
\end{eqnarray}
By making use of these solutions and assuming the on-shell condition $\Omega=\sqrt{k_y^2+k_z^2}$ for the emitted photons, the imaginary part of the (Lorentz-contracted) polarization tensor can be written as follows~\cite{Wang-Shovkovy:2020}:
\begin{eqnarray}
\mbox{Im} \left[\Pi^{\mu}_{R,\mu}\right] &=&
\sum_{f=u, d} \frac{N_c\alpha_f}{2\pi  l_f^4} \sum_{n>n^\prime}^{\infty}  
\frac{g(n, n^{\prime}) 
\left[
\Theta\left(k_{-}^{f}-|k_y|\right)
-\Theta\left(|k_y|-k_{+}^{f}\right) \right]
 }{\sqrt{ [( k_{-}^{f} )^2 - k_y^2 ][ (k_{+}^{f})^2- k_y^2] } } \left(
\mathcal{F}_1^f+\mathcal{F}_4^f \right)
\nonumber\\
&-&\sum_{f=u, d} \frac{N_c\alpha_f}{4\pi  l_f^4} \sum_{n=0}^{\infty} 
\frac{g_0(n)\Theta\left(|k_y|-k_{+}^{f}\right)}{\sqrt{ k_y^2 [ k_y^2-(k_{+}^{f})^2]} }\left(
\mathcal{F}_1^f+\mathcal{F}_4^f \right)  ,
\label{Im-Pi-final}
\end{eqnarray}
where $\Theta\left(x\right)$ is the Heaviside step function. We also used the following shorthand notations:
\begin{eqnarray}
g(n, n^{\prime}) &=& 2-\sum_{s_1,s_2=\pm}
n_F\left(\frac{\Omega}{2}  +s_1 \frac{\Omega(n-n^{\prime})|e_fB|}{k_y^2}+s_2 \frac{|k_z|}{2k_y^2}\sqrt{ \left(k_y^2-(k_{-}^{f})^2 \right)\left( k_y^2- (k_{+}^{f})^2\right)} \right), 
\label{gnn}  \\
g_0(n) &=& g(n, n)  =2 - 2\sum_{s=\pm}
n_F\left(\frac{\Omega}{2} +s \frac{|k_z|}{2|k_y|}\sqrt{ k_y^2 - 4(m^2+2n|e_f B|)}
\right) .
\label{g0n}
\end{eqnarray}
In the final expression, we took into account that the quark ($q\to q +\gamma$) and antiquark ($ \bar{q}\to \bar{q} +\gamma$) splitting processes contribute equally. Note that, in the problem at hand, this is the consequence of the charge-conjugation symmetry.

\section{Photon emission rates}
\label{sec:Numerical-results}

In this section, by making use of the explicit expression for the imaginary part of polarization function in Eq.~(\ref{Im-Pi-final}), we study numerically the photon emission rate in a strongly magnetized quark-gluon plasma. The main question that we aim to address is the dependence of emission on the magnitude and direction of the photon momentum. 

To optimize numerical calculations, we express all dimensionful quantities in units of the (neutral) pion mass, $m_{\pi}\approx 0.135~\mbox{GeV}$. This is indeed convenient since the corresponding energy scale is representative of the key properties of a quark-gluon plasma produced in heavy-ion collisions. This is also suitable for the purposes of this study since the magnetic field and temperature are of the order of the pion mass squared and the pion mass, i.e., $|eB|\sim m_\pi^2$ and $T \sim m_{\pi}$, respectively. 

To study the photon emission rate as a function of the transverse momentum $k_T$ (which is same as $\Omega$ here) and the azimuthal angle $\phi$, see Fig.~\ref{illustration}, we will use the parametrization for the photon momenta in Eq.~(\ref{kT-phi-parametrization}). For a better understanding of the photon emission, we will investigate in detail the cases of two representative choices of the magnetic field strength, $|eB|=m_\pi^2$ and $|eB|=5m_\pi^2$, and two representative values of temperature, $T=0.2~\mbox{GeV}$ and $T=0.35~\mbox{GeV}$. In each case, we will limit the range of photon parameters as follows. The transverse momenta will be taken in the range between $k_{T,min}=0.01~\mbox{GeV}$ and $k_{T,max}=1~\mbox{GeV}$, with the discretization step $\Delta k_{T}=0.01~\mbox{GeV}$. The symmetry of the problem implies that the emission should be invariant with respect to a mirror reflection in the reaction plane (i.e., $\phi \to -\phi$). Thus, it is sufficient for us to cover the range of azimuthal angles between $\phi=0$ and $\phi=\frac{\pi}{2}$. In the actual numerical calculations, however, we will be avoiding the limiting values of the azimuthal angle by considering the range between $\phi_{min}=10^{-4}\frac{\pi}{2}$ and $\phi_{max}=\frac{\pi}{2} - \phi_{min}$ and use the discretization step $\Delta \phi=10^{-3}\frac{\pi}{2}$. 

When calculating numerically the sum over Landau levels in Eq.~(\ref{Im-Pi-final}), we will approximate the result by including only a finite number of terms with $n,n^{\prime}\leq n_{max}$. Qualitatively, this corresponds to setting an ultraviolet energy cutoff at $\Lambda\simeq \sqrt{2n_{max} |eB|}$. Of course, the minimum value of $n_{max}$ should be sufficiently large to include all Landau levels that contribute substantially to the photon rate. From the kinematics of the relevant $1\to 2$ and $2\to 1$ processes shown in Fig.~\ref{LLprocesses}, we find that the most stringent constraints come from the regions of very small and very large transverse momenta. They give $n_{max}\gtrsim |eB|/k_{T, min}^{2}$ and $n_{max}\gtrsim k_{T, max}^{2}/|eB|$, respectively. The first constraint (i.e., $n_{max}\gtrsim |eB|/k_{T, min}^{2}$) is the consequence of the Landau-level quantization. As is clear, the emission of photons with a small $k_{T}$ (energy) is possible only when the energy separation between the nearest Landau levels is smaller than $k_{T}$. As we explain in detail later, this is true only for the quark states with energies of the order of $\sim |eB|/k_T$ or larger. Thus, to include all relevant states with such high energies, one needs to sum up at least $n_{max}\gtrsim |eB|/k_{T, min}^{2}$ Landau levels. The other constraint (i.e., $n_{max}\gtrsim k_{T, max}^{2}/|eB|$) comes from the requirement of having a nonempty phase space when the value of $k_{T, max}$ (energy) is large. This is also easy to understand since the emission of high-energy photons would be impossible if the quark states with sufficiently large energies were removed from the spectrum.

By choosing $k_{T,min}=0.01~\mbox{GeV}$ and $k_{T,max}=1~\mbox{GeV}$, one finds that the minimum number of $n_{max}$ should be greater than $\mbox{max}(182,55)$ when $|eB| =m_\pi^2$. Similarly, in the case of $|eB| =5m_\pi^2$, one finds that $n_{max}$ should be greater than $\mbox{max}(911,11)$. Therefore, to cover both choices of the magnetic field and the whole range of transverse momenta, we will use a relatively large number of Landau levels, i.e., $n_{max} =1000$. Numerically, the latter corresponds to the ultraviolet energy cutoff $\Lambda\simeq 6~\mbox{GeV}$ for $|eB|=m_\pi^2$  and $\Lambda\simeq13.5~\mbox{GeV}$ for $|eB|=5m_\pi^2$. With limited computational recourses, however, the maximum number of Landau levels could be chosen on the case-by-case basis. In particular, this could lead to a substantial reduction of $n_{max}$ when the photon transverse momenta are neither too large nor too small. It should be mentioned that the above consideration is valid only in the regime when the magnetic energy scale $\sqrt{|eB|}$ is comparable to temperature. At temperatures much higher or much lower than $\sqrt{|eB|}$, the estimates for $n_{max}$ will be different.

\begin{figure}[t]
\centering
\subfigure[]{\includegraphics[width=0.45\textwidth]{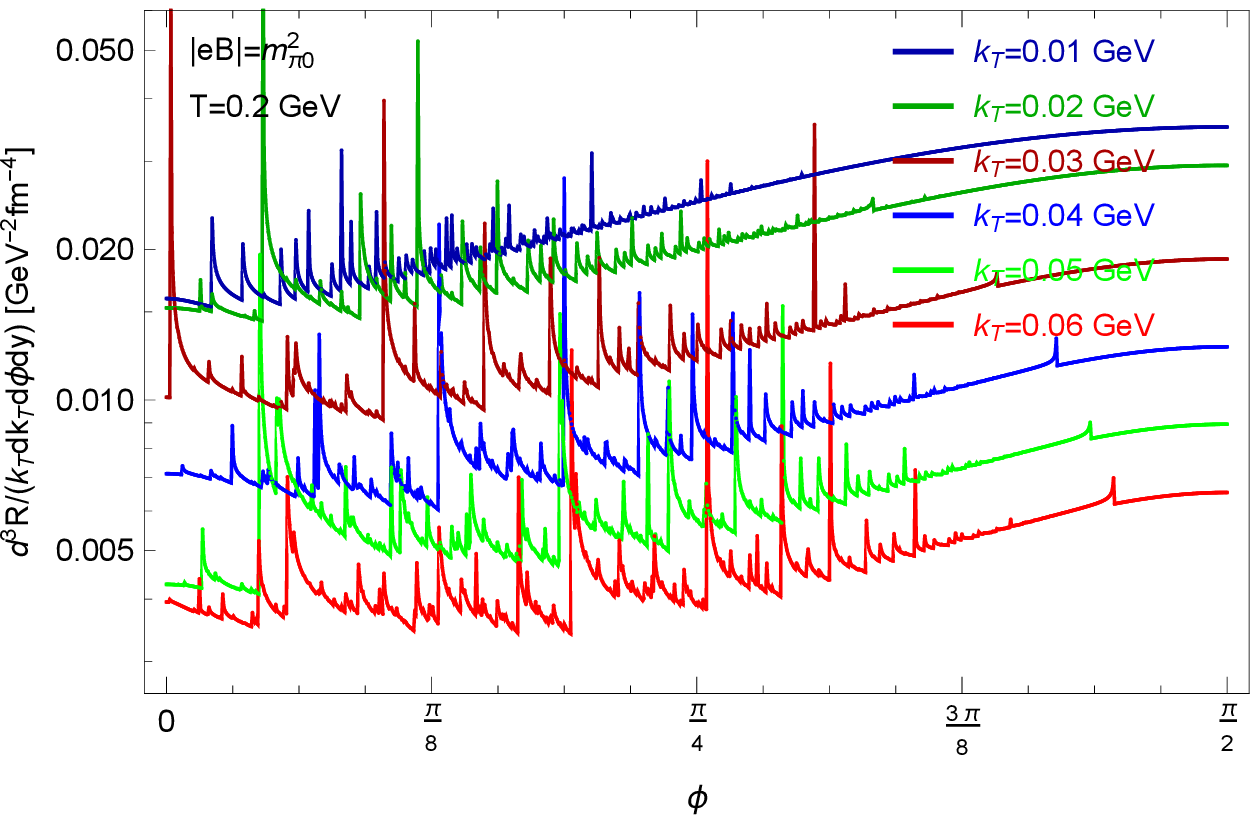}}
  \hspace{0.01\textwidth}
\subfigure[]{\includegraphics[width=0.45\textwidth]{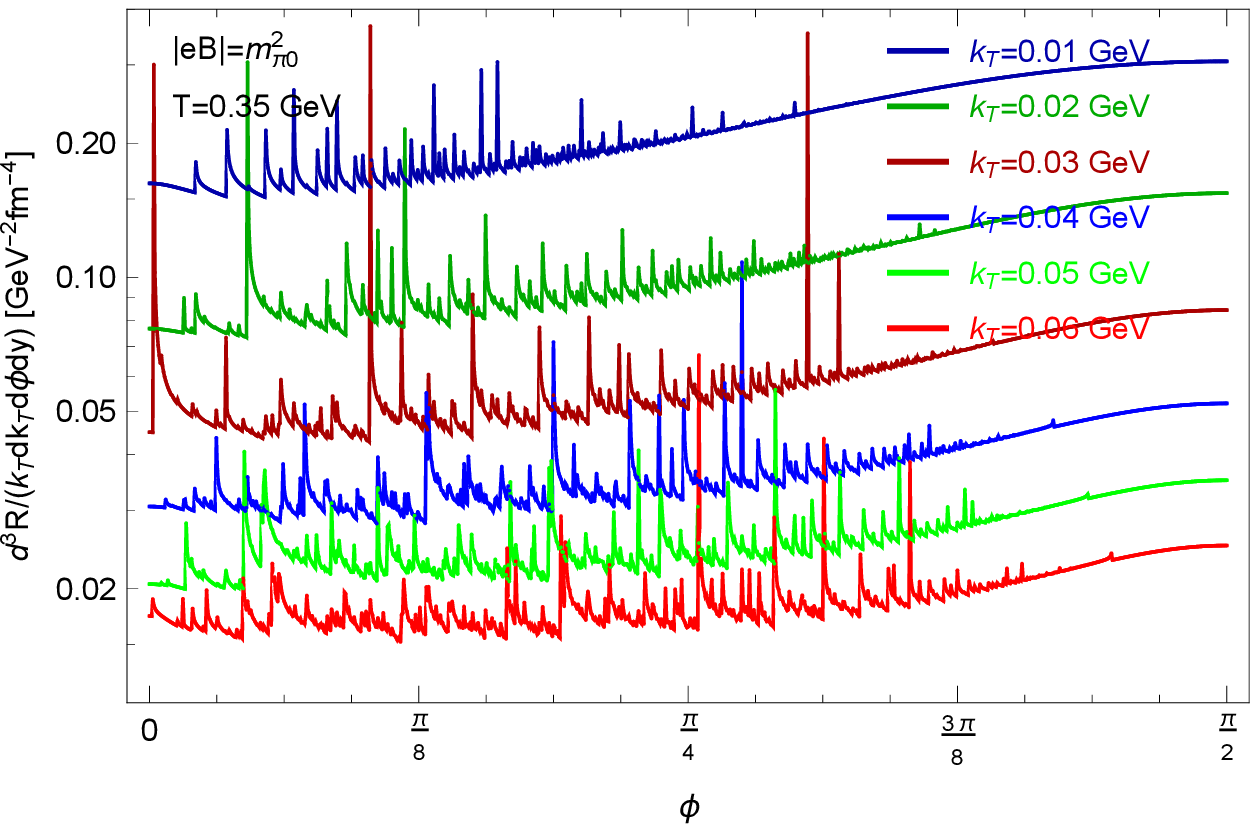}}\\
\subfigure[]{\includegraphics[width=0.45\textwidth]{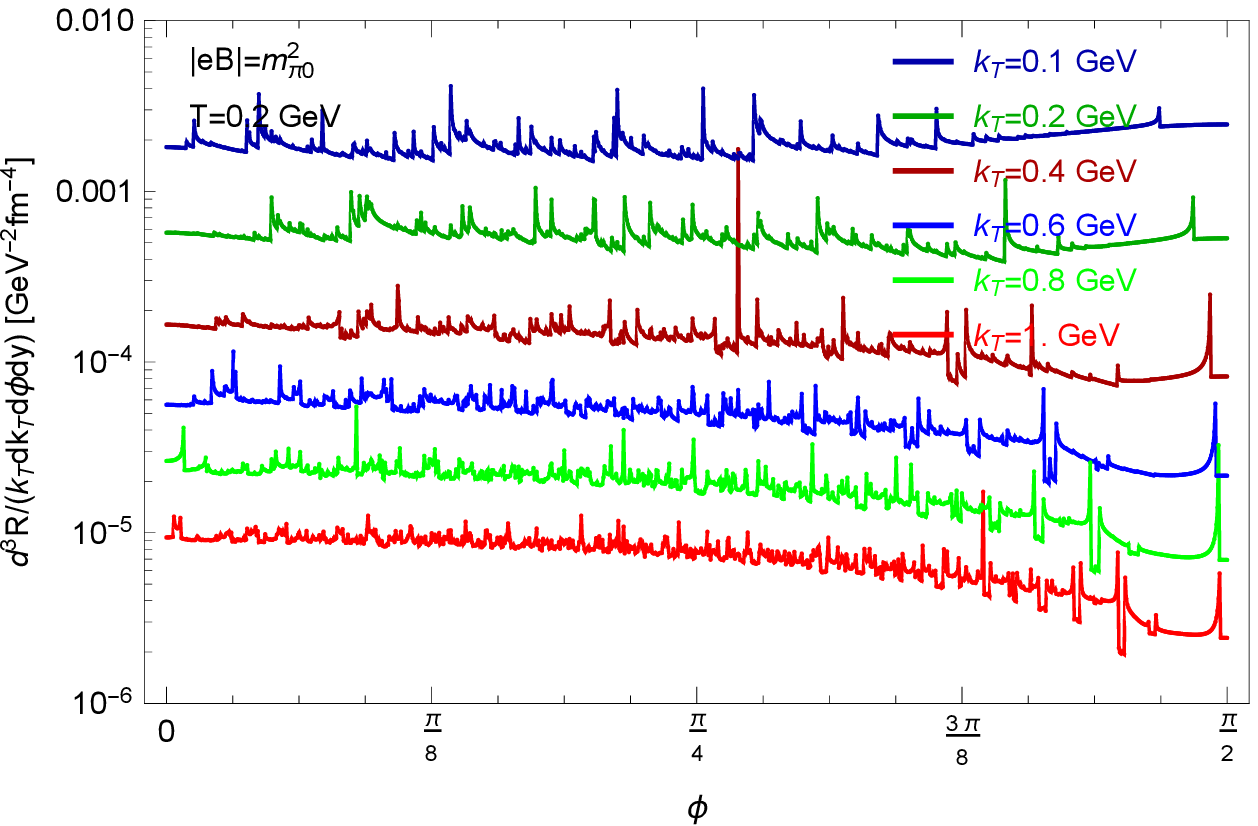}}
  \hspace{0.01\textwidth}
\subfigure[]{\includegraphics[width=0.45\textwidth]{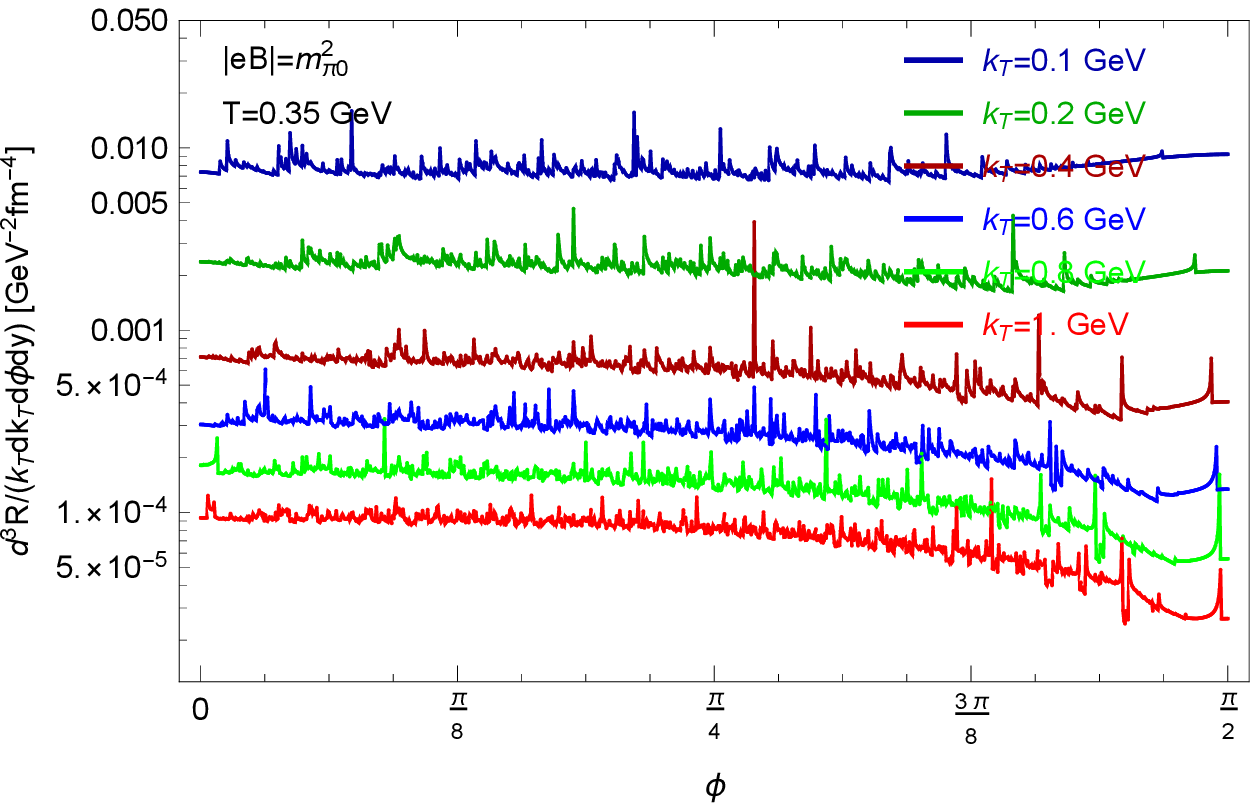}}
\caption{The angular dependence of the photon production rates for $|eB|=m_\pi^2$ 
and fixed values of $k_T$. The left panels show the results for $T=200~\mbox{MeV}$ 
(panels a and c) and the right panels for $T=350~\mbox{MeV}$ (panels b and d).
The top (bottom) panels show the results for several small (large) values of $k_T$.}
\label{fig:B1:rates}
\end{figure}

Let us start by considering the case of moderately strong magnetic field, $|eB|=m_\pi^2$. The corresponding numerical results for the angular dependence of the photon emission rates are shown in Fig.~\ref{fig:B1:rates} for selected values of the transverse momentum $k_T$. The two panels on the left (a and c) give the rates for $T=0.2~\mbox{GeV}$ and the two panels on the right (b and d)  for $T=0.35~\mbox{GeV}$. The two upper (lower) panels correspond to a range of small (large) values of $k_T$. As is clear form the figure, there are several qualitative features of the photon emission that stand out. 

First, as we see, the rate is not a smooth function of $\phi$. It is easy to understand that this is the consequence of the Landau level quantization for quark states that causes numerous threshold effects in the photon production. In principle, the corresponding threshold effects should be smoothed out by a nonzero quasiparticle width of quarks due to their interactions in plasma. Conceptually, the quasiparticle width can be obtained from the imaginary part of the quark self-energy. In the case of hot QCD, of course, the latter should be dominated by the gluon-exchange interaction. While the interaction effects are expected to smooth out the angular dependence (as well as its energy dependence), they are not expected to change qualitatively the overall features in the photon production. This is indeed reasonable since the combined effect of numerous Landau levels will largely average in a plasma with temperature $T\gtrsim \sqrt{|eB|}$. Thus, for simplicity, we will neglect the gluon-mediated interaction effects and assume that quarks have the vanishing quasiparticle width in this study. 

By comparing the results for different values of $k_T$ in Fig.~\ref{fig:B1:rates}, we see that, on average, the rate tends to decrease with increasing of the transverse photon momentum (or, equivalently, the energy). This is explained in part by the suppression of all processes involving quarks and antiquarks with large energies. Mathematically, this comes from the Fermi-Dirac distribution functions in the polarization tensor, see Eqs.~(\ref{Im-Pi-final}) -- (\ref{g0n}). The other contributing factor to the suppression at large   transverse momenta is the overall Bose distribution factor $1/[\exp\left(\Omega/T\right)-1]$ in definition of the rate in Eq.~(\ref{diff-rate-2}). 

By carefully analyzing the contributions of different types of processes, we find that the photon production rate is largely dominated by the two splitting processes $q\rightarrow q+\gamma $ and $\bar{q} \rightarrow \bar{q}+\gamma $ for a wide range of moderately high temperatures ($T\gtrsim m_\pi$), moderately strong magnetic fields ($|eB|\gtrsim m_\pi^2$), and not too larger transverse momenta ($k_T \lesssim \sqrt{|eB|}$). With increasing $k_T$, however, the relative contribution of the annihilation process $q + \bar{q}\rightarrow \gamma$ grows gradually. From our numerical results, we find that it gives a comparable contribution when $k_T \gtrsim 0.5~\mbox{GeV}$ or so. (One can also verify that the annihilation process plays the dominant role at very small temperatures, $T \ll \sqrt{|eB|}$, but such a regime of quark-gluon plasma is irrelevant in the context of heavy-ion collisions.)

As we see from panels (a) and (b) in Fig.~\ref{fig:B1:rates}, the emission rate at small values of $k_T$ has an overall tendency to peak at $\phi = \frac{\pi}{2}$, i.e., in the direction perpendicular to the reaction plane. This behavior changes dramatically at large values of $k_T$, as seen from panels (c) and (d) in Fig.~\ref{fig:B1:rates}. Indeed, when the value of $k_T$ is larger than about $\sqrt{|eB|}$, the emission tends to be highest at $\phi = 0$, i.e., in the direction along the reaction plane. Such unusual behavior has interesting underlying physics and may have important implications. In application to heavy-ion collisions, for example, one can argue that the direct photon production will be characterized by an apparent flow with a negative ellipticity coefficient $v_2$ at small values of $k_T$ and a positive $v_2$ at large values of $k_T$. Of course, such flow is caused by a strong magnetic field and has nothing to do with the hydrodynamic behavior of the quark-gluon plasma. 

\begin{figure}[t]
\centering
\subfigure[]{\includegraphics[width=0.45\textwidth]{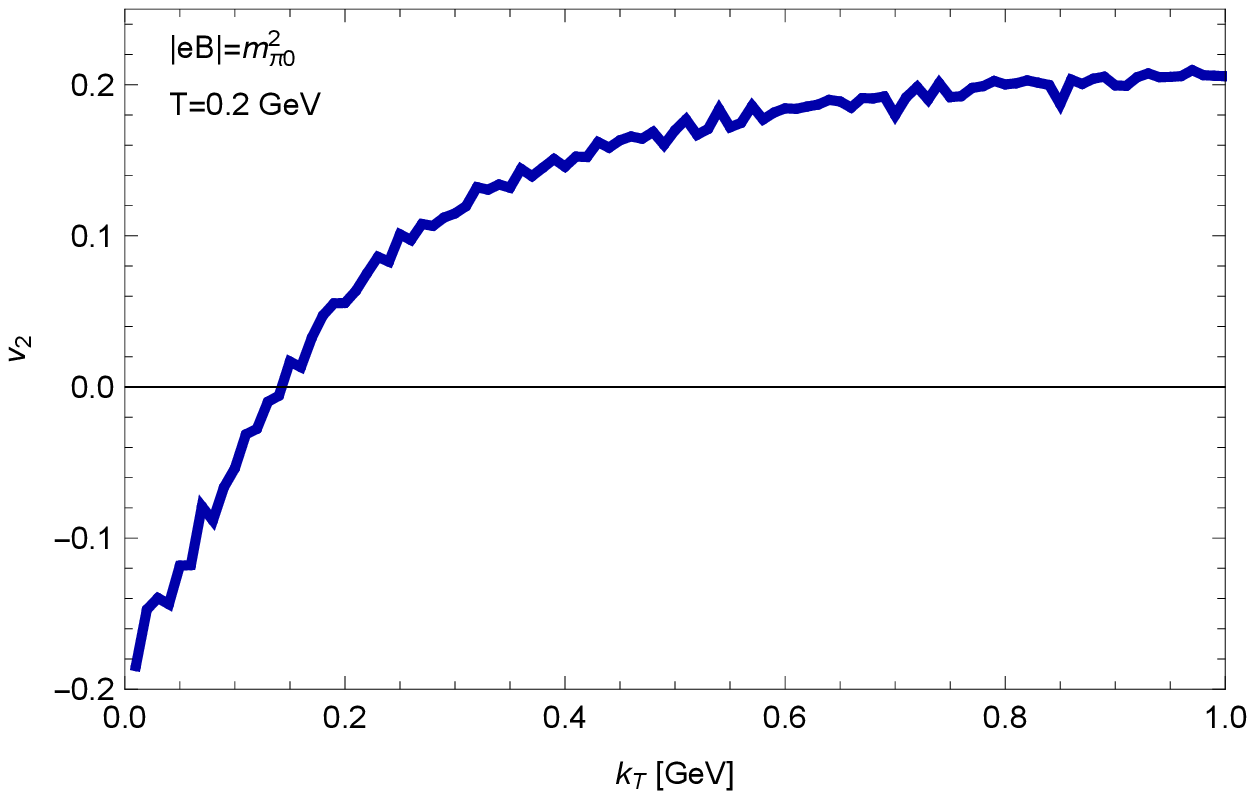}}
  \hspace{0.01\textwidth}
\subfigure[]{\includegraphics[width=0.45\textwidth]{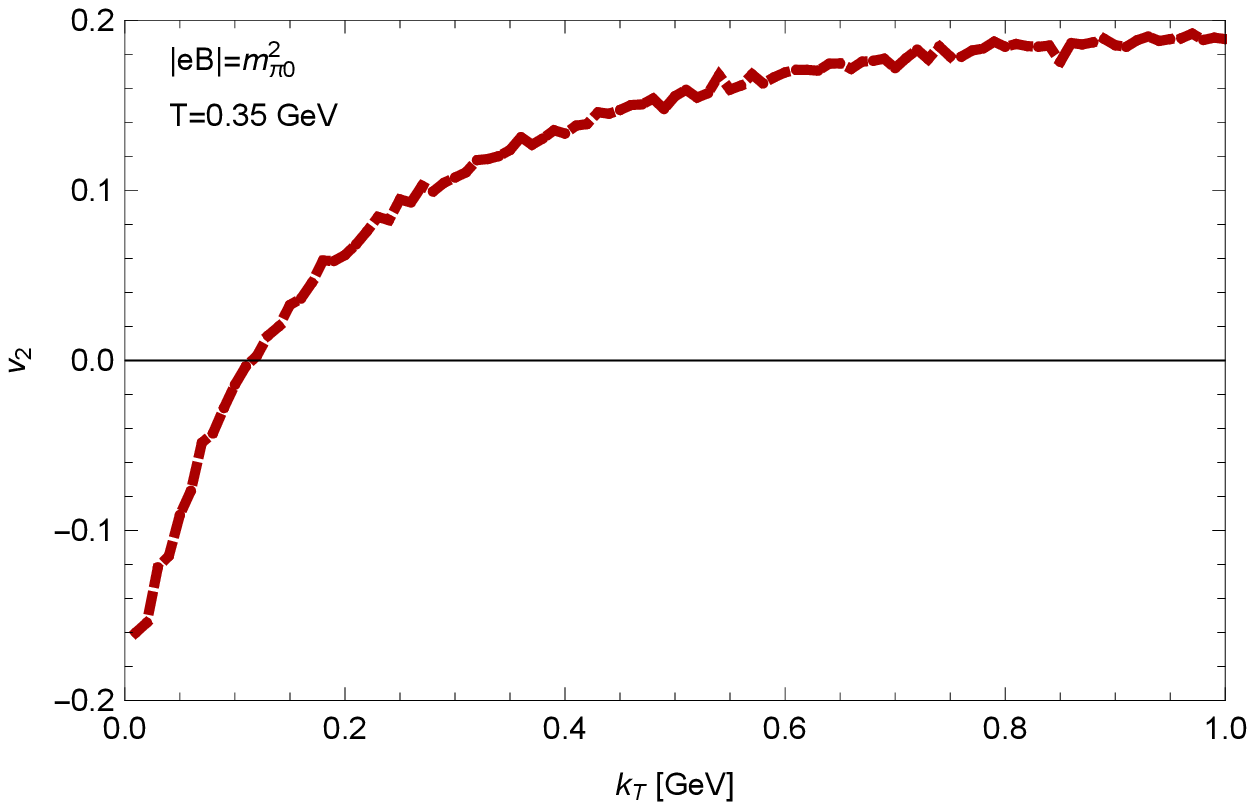}}
\caption{Ellipticity of the photon production as a function of the transverse momentum $k_T$ for $|eB|=m_\pi^2$ and 
two different temperatures: $T=200~\mbox{MeV}$ (panel a) and $T=350~\mbox{MeV}$ (panel b).}
\label{fig:B1:v2}
\end{figure}

The qualitative finding about the apparent flow of the direct photon emission can be formally verified by calculating the ellipticity coefficient $v_2$ by using the definition in Eq.~(\ref{v2}). Our numerical results for $v_2$ as a function of $k_T$ are presented  in Fig.~\ref{fig:B1:v2} for the same two values of temperature, i.e., $T=0.2~\mbox{GeV}$ (panel a) and $T=0.35~\mbox{GeV}$ (panel b). As anticipated, the ellipticity coefficient $v_2$ takes negative values at small $k_T$ and positive values at large $k_T$. The critical point where $v_2$ vanishes appears to be roughly around $k_T \simeq \sqrt{|eB|}$. Needless to say, because of the quantization of Landau levels and numerous threshold effects, the functional dependence of $v_2$ on $k_T$ is not a smooth function. However, there is a clear tendency of $v_2$ to grow with $k_T$. Despite the large difference in the overall photon production rate at two different temperatures, our results in Fig.~\ref{fig:B1:v2} do not reveal a strong dependence of $v_2$ on temperature. The only exception is, perhaps, the region of small transverse momenta, where larger negative values of $v_2$ can be achieved with decreasing temperature.  

One of the most interesting features of the $v_2$ dependence on the transverse momentum is its behavior at large $k_T$. As is clear from our calculation, the ellipticity reaches and saturates at a relatively large positive value, i.e, $v_{2,max} \simeq 0.2$. From a physics viewpoint, one might wonder why $v_2$ does not vanish when the  transverse momentum is much larger than the magnetic energy scale $\sqrt{|eB|}$. The reason is quite simple and is connected with the underlying mechanism of the photon emission in a magnetized plasma. In essence, it is the magnetic field in the first place that makes the corresponding photon emission possible without the mediation of any additional particles (e.g., gluons) in the initial or final states. Thus, while the total integrated photon rate quickly decreases with increasing $k_T$, the angular dependence preserves a characteristic oblate shape described by a moderately large positive $v_2$. It is appropriate to mention that the magnitude of $v_{2,max} \simeq 0.2$ appears to be considerably smaller than $4/7\approx 0.57$ predicted by the classical analysis in Ref.~\cite{Tuchin:2014pka}. We assume that this is largely due to the inclusion of the annihilation processes in our analysis.

It is tempting to argue that the predicted positive $v_2$ at large $k_T$ could be very important in the context of heavy-ion collisions, where the direct photon production is characterized by a surprisingly large $v_2$. Of course, a more realistic and complete model of the direct photon production form a hot quark-gluon plasma should include not only the emission assisted by a magnetic field but also the gluon-mediated processes \cite{Kapusta:1991qp,Baier:1991em,Aurenche:1998nw,Steffen:2001pv,Arnold:2001ba,Arnold:2001ms,Ghiglieri:2013gia}. The latter are expected to be nearly isotropic in the local rest frame. It is the relative weight of the gluon and magnetic field mediated processes that should determine the net $v_2$. It should be pointed that, at large values of $k_T$, the total integrated rates for both types of processes are strongly suppressed by the Fermi-Dirac distributions of quarks. Unlike the gluon-mediated processes, which are additionally suppressed by the Bose distributions of gluons, the leading order splitting and annihilation processes in a magnetic field do not suffer from an extra suppression. There is a good chance, therefore, that the latter play an important role indeed. A careful investigation of this issue deserves a separate study, however. 

\begin{figure}[t]
\centering
\subfigure[]{\includegraphics[width=0.45\textwidth]{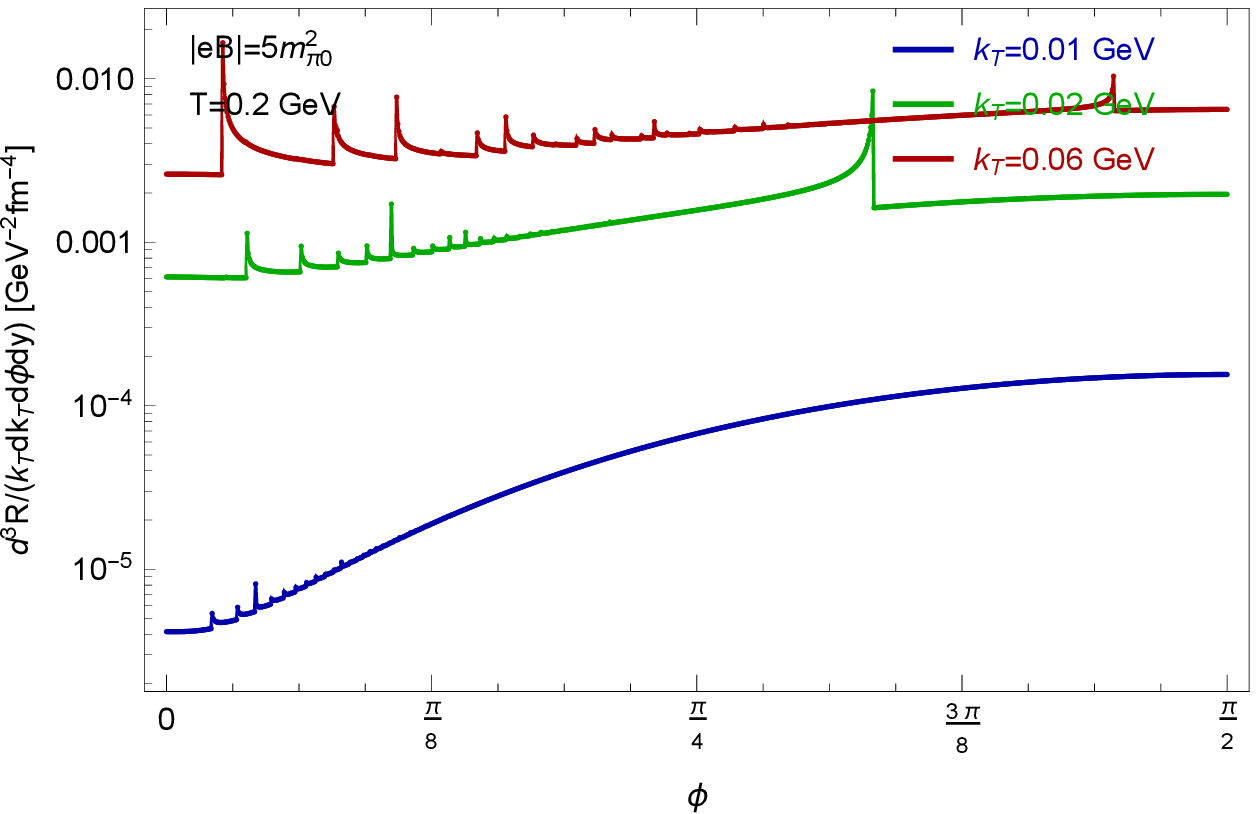}}
  \hspace{0.01\textwidth}
\subfigure[]{\includegraphics[width=0.45\textwidth]{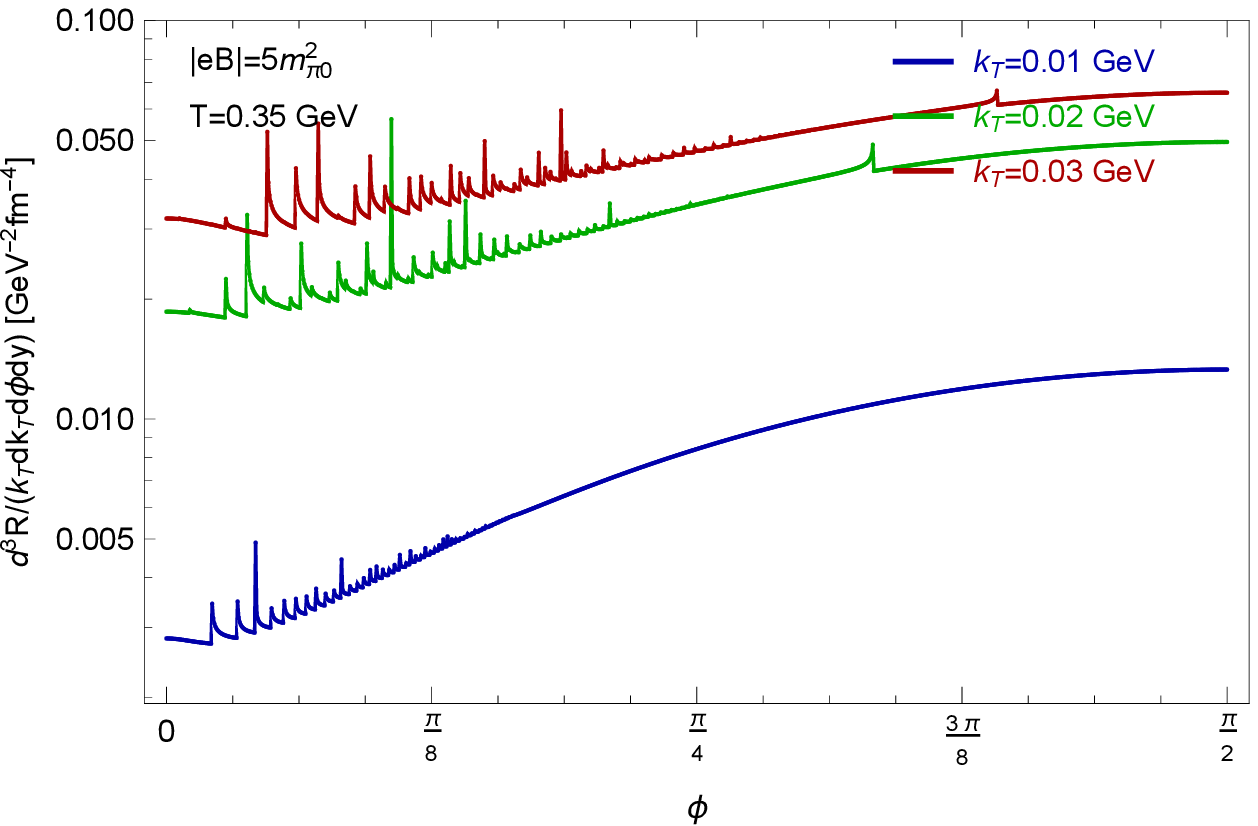}}\\
\subfigure[]{\includegraphics[width=0.45\textwidth]{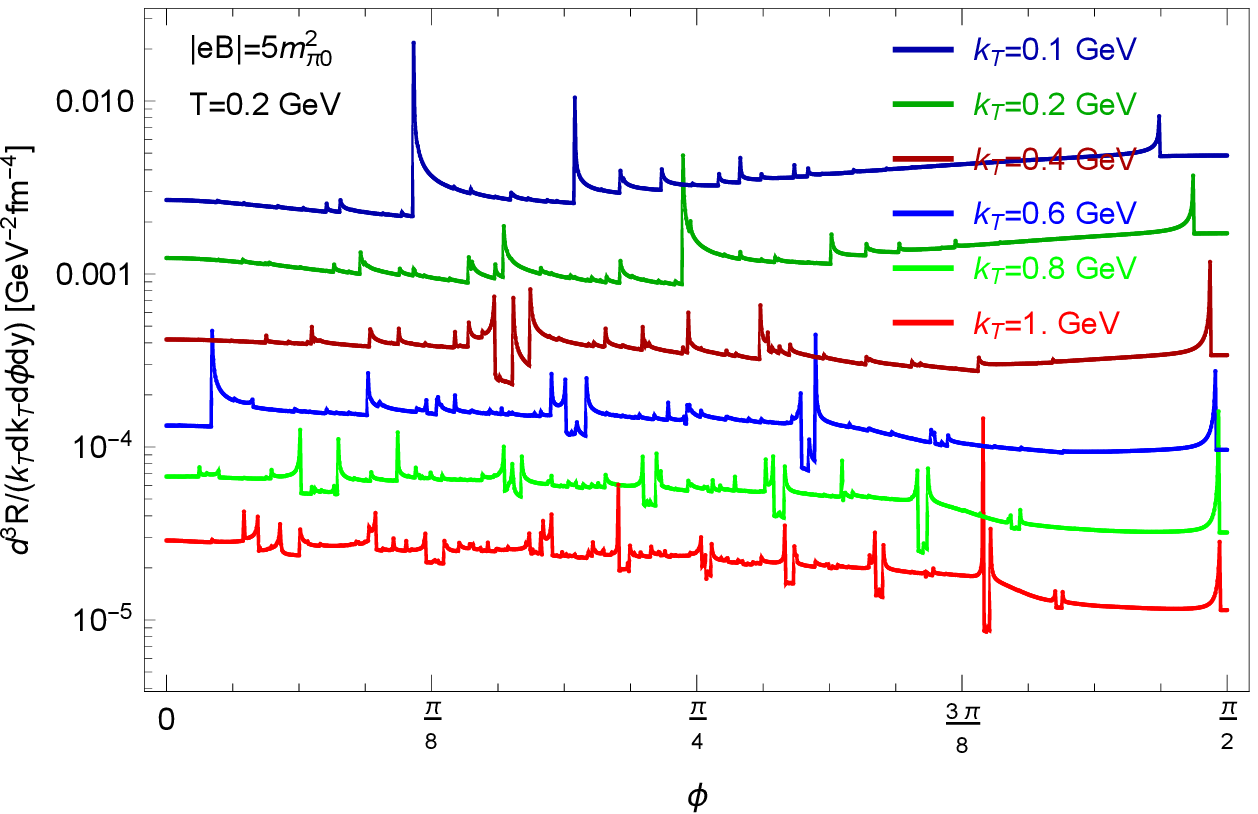}}
  \hspace{0.01\textwidth}
\subfigure[]{\includegraphics[width=0.45\textwidth]{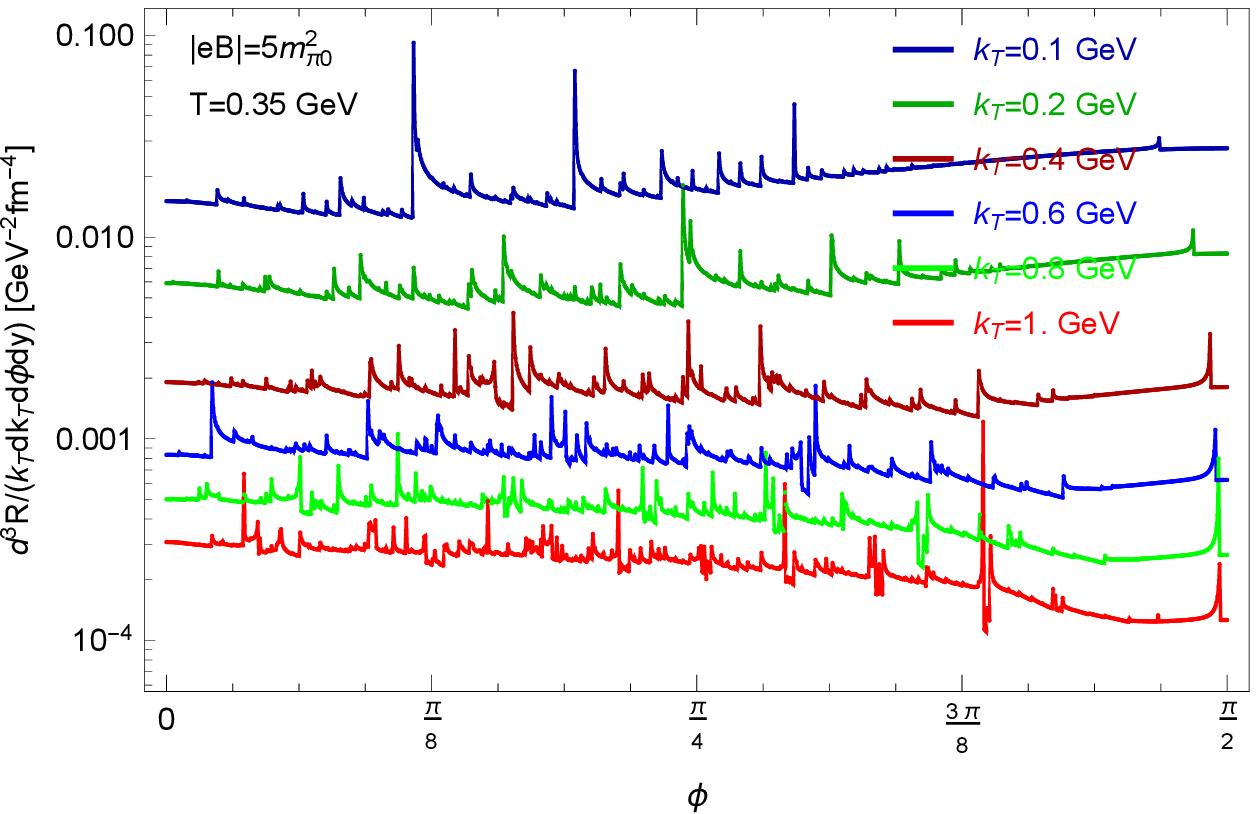}}
\caption{The angular dependence of the photon production rates for $|eB|=5 m_\pi^2$ 
and fixed values of $k_T$. The left panels show the results for $T=200~\mbox{MeV}$ 
(panels a and c) and the right panels for $T=350~\mbox{MeV}$ (panels b and d).
The top (bottom) panels show the results for several small (large) values of $k_T$.}
\label{fig:B5:rates}
\end{figure}

\begin{figure}[t]
\centering
\subfigure[]{\includegraphics[width=0.45\textwidth]{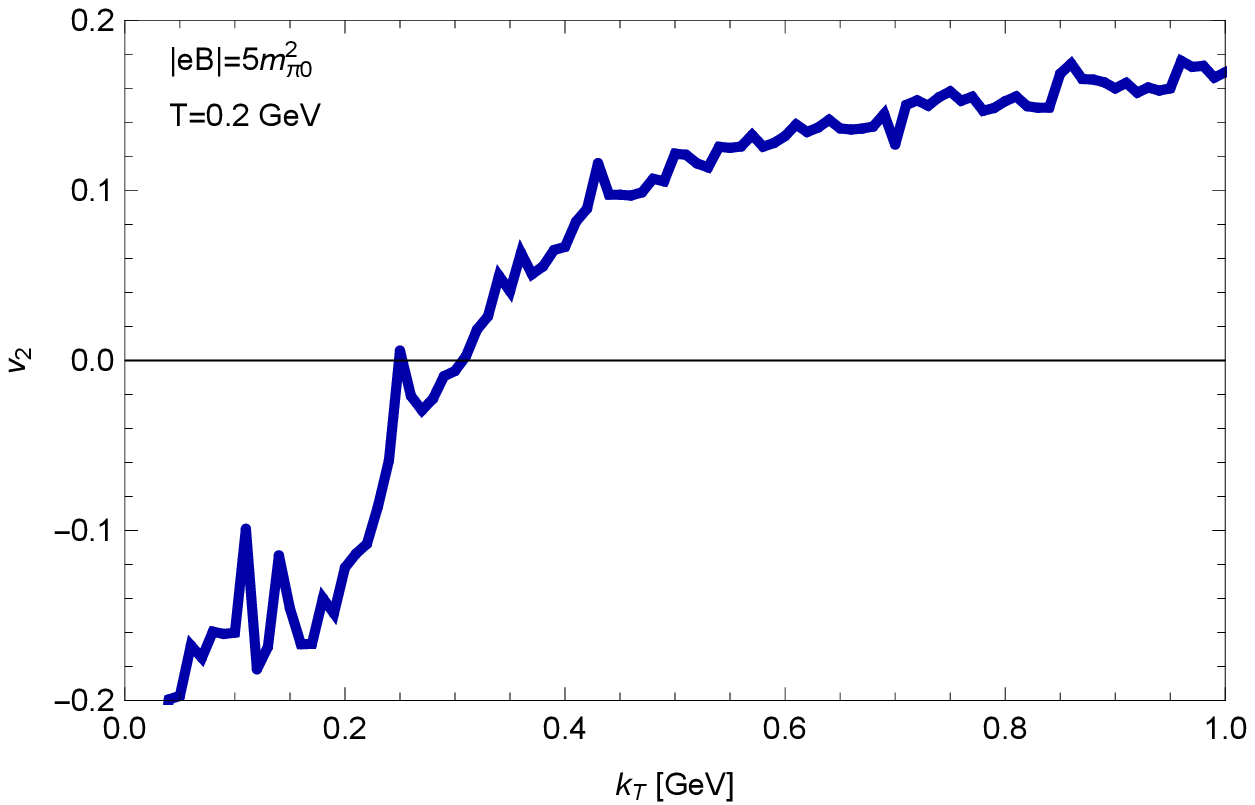}}
  \hspace{0.01\textwidth}
\subfigure[]{\includegraphics[width=0.45\textwidth]{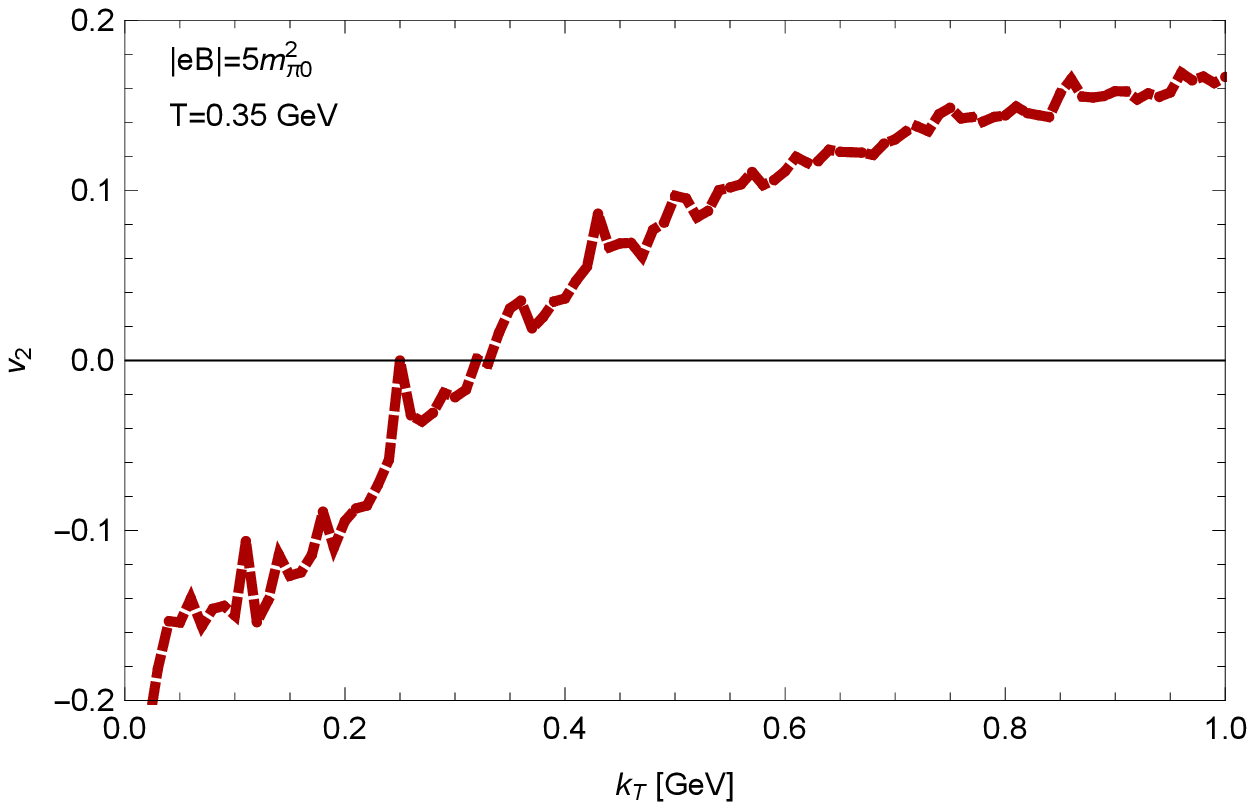}}
\caption{Ellipticity of the photon production as a function of the transverse momentum $k_T$ for $|eB|=5m_\pi^2$ and 
two different temperatures:  $T=200~\mbox{MeV}$ (panel a) and $T=350~\mbox{MeV}$ (panel b).}
\label{fig:B5:v2}
\end{figure}

In order to better understand the role of the magnetic field on the direct photon emission from hot quark-gluon plasma, it is instructive to consider the case of a stronger field. So, let us now consider the case with $|eB|=5 m_\pi^2$. The corresponding angular dependence of the photon emission rates are shown in Fig.~\ref{fig:B5:rates} for a wide range of values of the transverse momenta. As in the case of the weaker field, we present the results for $T=0.2~\mbox{GeV}$ in the two left panels (a and c) and $T=0.35~\mbox{GeV}$ in the two right panels (b and d). The pair of upper (lower) panels show the results for small (large) values of $k_T$. The corresponding results for the ellipticity of the emission are shown in Fig.~\ref{fig:B5:v2}. Needless to say that the $v_2$ dependence on the transverse momentum is qualitatively the same as in the case of a weaker field. It stays negative at small $k_T$, crosses zero around $k_T \simeq \sqrt{|eB|}$, and then remains positive at large $k_T$, with the saturation value again close to $v_{2,max}\simeq 0.2$. 

While the results for a stronger magnetic field, $|eB|=5 m_\pi^2$, shares many similarities with that of a weaker field, $|eB|=m_\pi^2$, some qualitative differences are seen too. To start with, let us point that the photon rate tends to grow (rather than fall) with $k_T$ in a window of small values of $k_T$, see panels (a) and (b) in Fig.~\ref{fig:B5:rates}. From the available data, we determine that the maximum is reached at about $k_T\simeq 0.06~\mbox{GeV}$ when $T=0.2~\mbox{GeV}$ and at about $k_T\simeq 0.04~\mbox{GeV}$ when $T=0.35~\mbox{GeV}$. When the transverse momentum increases further, the photon production rate starts to decrease quickly, following the same qualitative behavior as seen before in the case of a weaker field.   

\begin{figure}[t]
\centering
\subfigure[]{\includegraphics[width=0.45\textwidth]{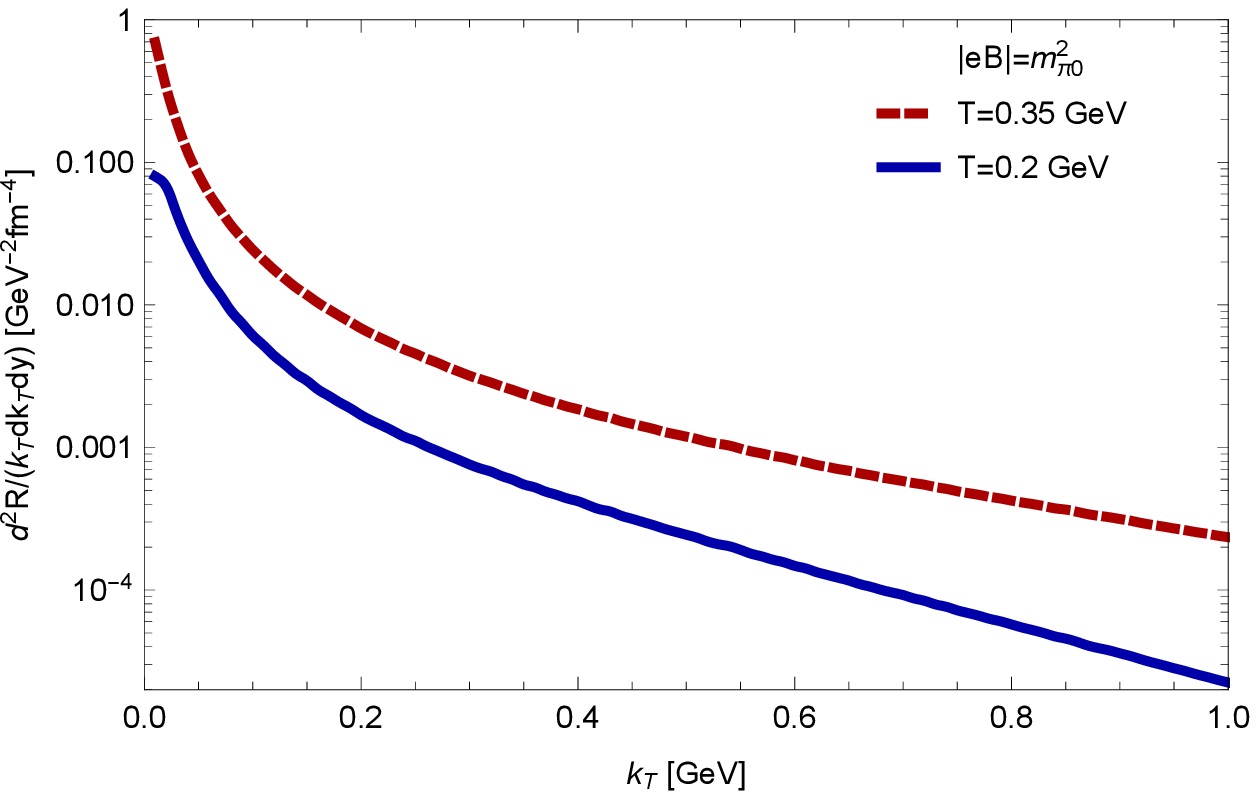}}
  \hspace{0.01\textwidth}
\subfigure[]{\includegraphics[width=0.45\textwidth]{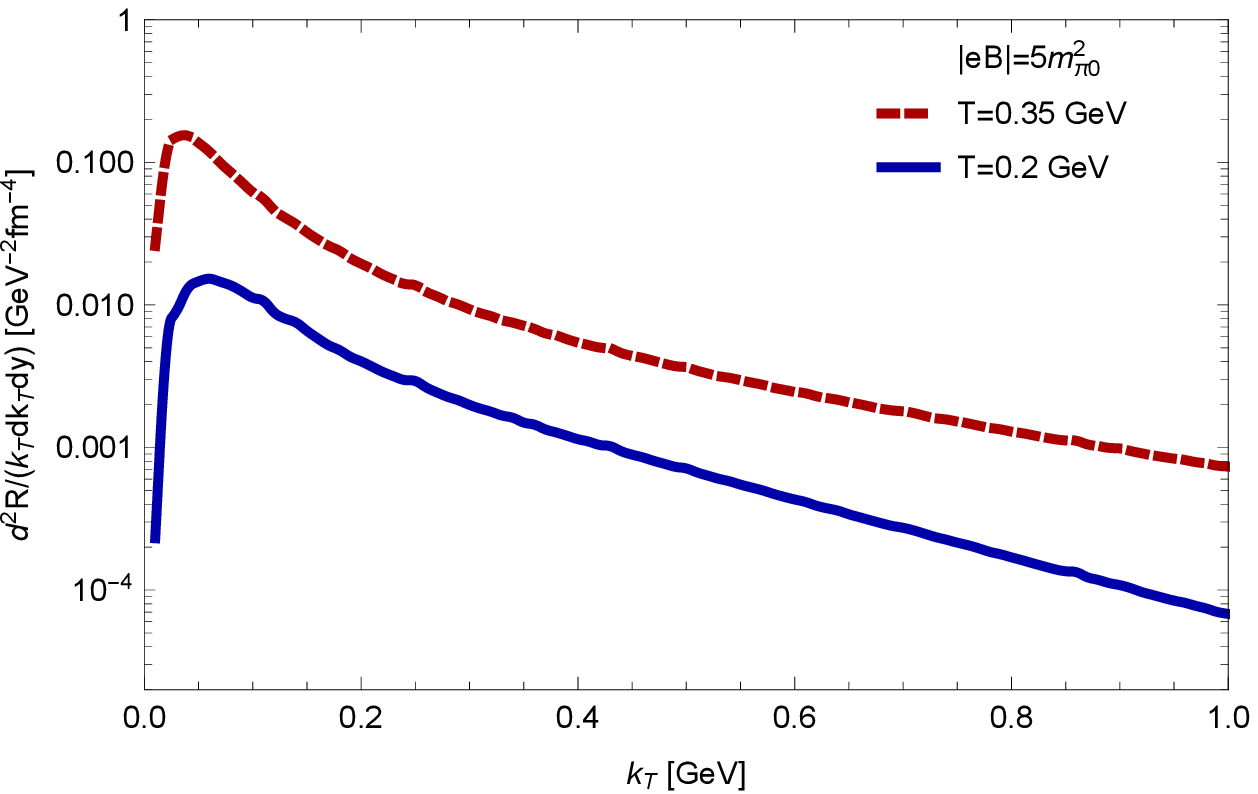}}
\caption{The integrated photon production rate as a function of the transverse momentum $k_T$ for $|eB|=m_\pi^2$ (panel a) and $|eB|=5m_\pi^2$ (panel b). The different lines represent results for two different temperatures, i.e., $T=200~\mbox{MeV}$ (blue solid line) and $T=350~\mbox{MeV}$ (red dashed line).}
\label{fig:ProdRateInt}
\end{figure}

In order to reconfirm the nonmonotonic dependence of the photon production rate on the transverse momentum, it is instructive to calculate the total rate integrated over the angular coordinate, as defined by Eq.~(\ref{Integrated-rate}). The corresponding results are plotted in Fig.~\ref{fig:ProdRateInt} for both choices of magnetic field, $|eB|=m_\pi^2$ (panel a) and $|eB|=5m_\pi^2$ (panel b). 
As clear from Fig.~\ref{fig:ProdRateInt}(b), in the case of the stronger field, $|eB|=5m_\pi^2$, there are indeed well-resolved peaks in the photon production rate at $k_T\simeq 0.06~\mbox{GeV}$ when $T=0.2~\mbox{GeV}$ and at $k_T\simeq 0.04~\mbox{GeV}$ when $T=0.35~\mbox{GeV}$. While similar peaks appear to be absent in the case of weaker field, $|eB|=m_\pi^2$, such a conclusion is premature. 

A careful analysis reveals that the photon rates must always have well-defined maxima at sufficiently small values of the transverse momentum. The existence of such maxima is a necessary consequence of the Landau-level quantization of quark states in a strongly magnetized plasma. In order to understand the underlying physics, it is instructive to consider in detail the kinematics of the quark splitting process $q\rightarrow q+\gamma $ in the regime of small $k_T$. Because of the charge-conjugation symmetry, the same is true for the antiquark splitting process $\bar{q} \rightarrow \bar{q}+\gamma $. As for the quark-antiquark  annihilation $q + \bar{q}\rightarrow \gamma$, it can be neglected in the regime of moderately high temperatures, $T\gtrsim m_\pi$, which is assumed here. 

\begin{figure}[t]
\centering
\subfigure[]{\includegraphics[width=0.45\textwidth]{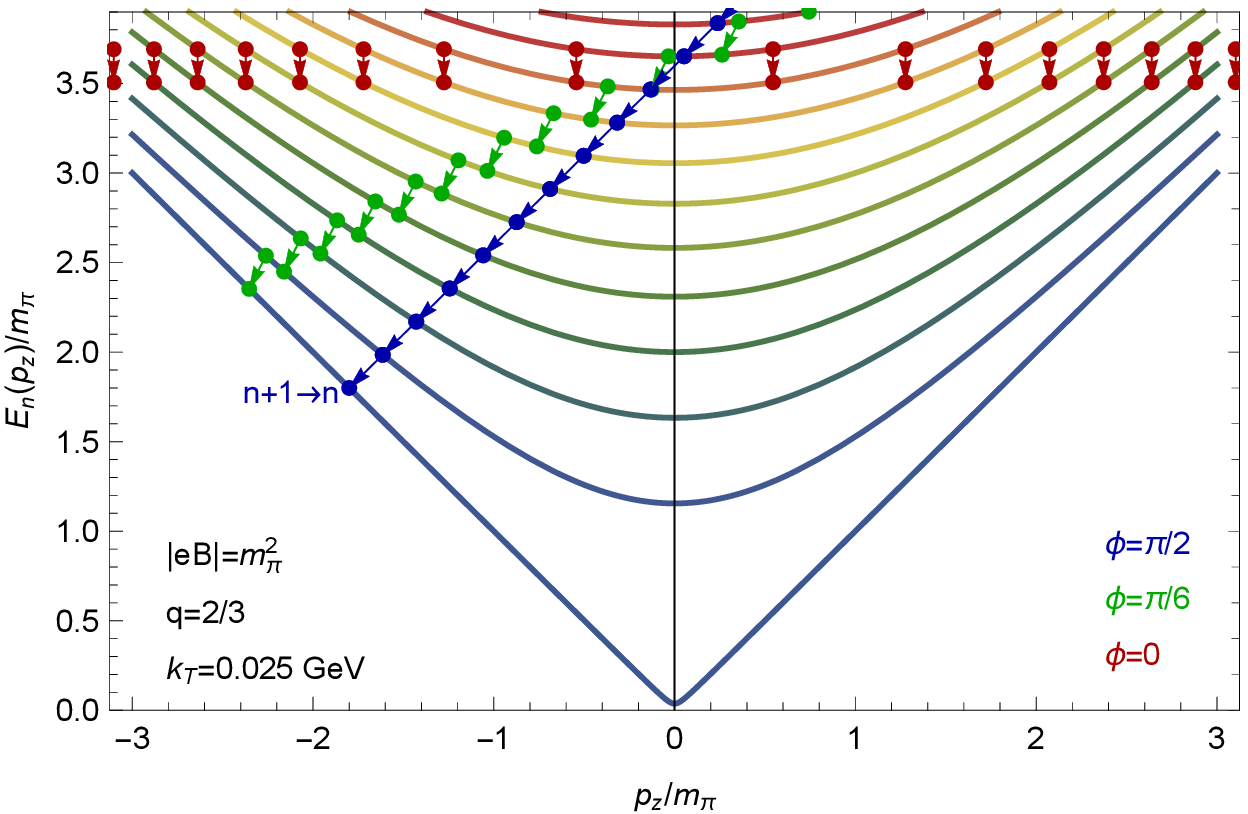}}
  \hspace{0.01\textwidth}
\subfigure[]{\includegraphics[width=0.45\textwidth]{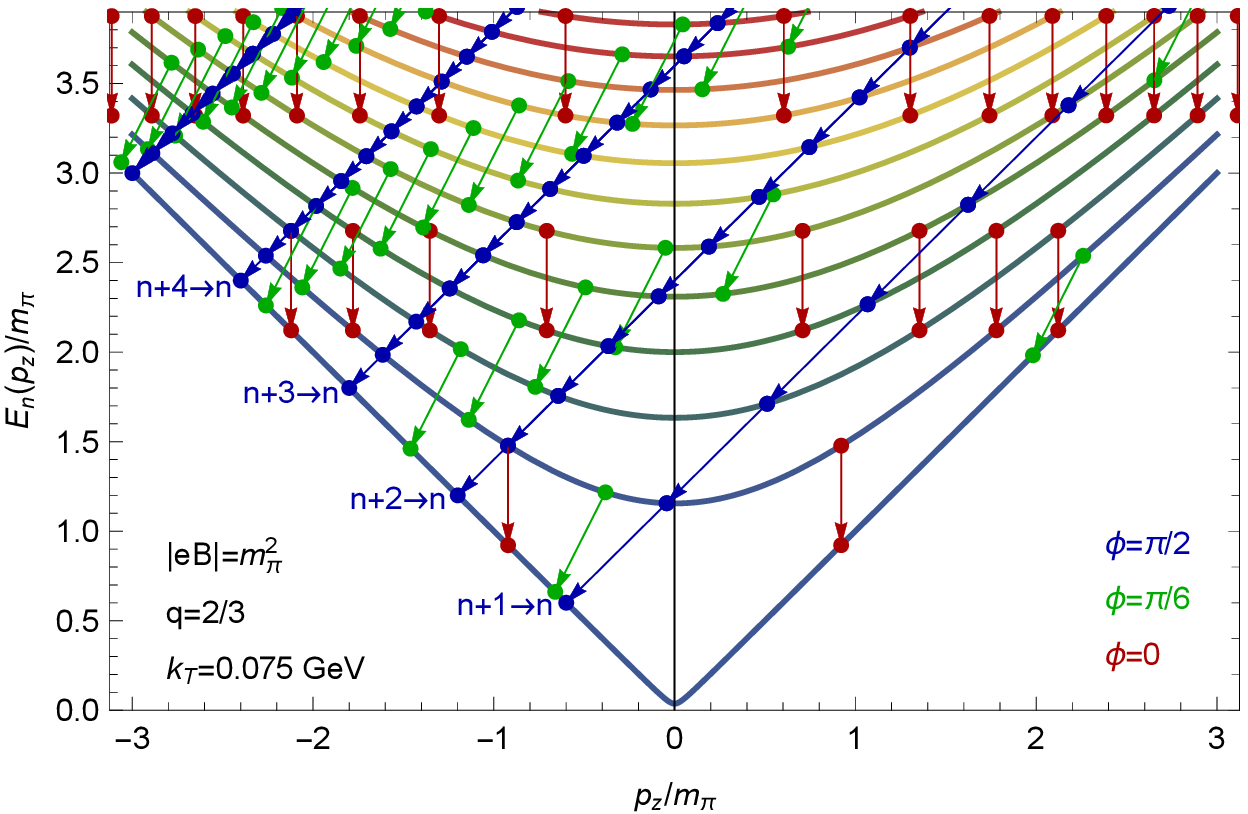}}
\caption{A schematic representation of the low-energy transitions between Landau levels 
that correspond to the quark splitting processes $q\to q+\gamma$, allowed by the 
energy conservation constraint. Transitions associated with the emission of photons in three 
different directions are color-coded as follows: $\phi=0$ (red), $\phi=\pi/6$ (green), and 
$\phi=\pi/2$ (blue). Note that only the transitions between the adjacent Landau levels 
($n+1\to n$) contribute nontrivially at $\phi = \pi/2$. The two choices of the photon 
transverse momenta are $k_T=0.025~\mbox{GeV}$ (panel a) and $k_T=0.075~\mbox{GeV}$ 
(panel b).}
\label{fig:transitions}
\end{figure}

The Landau level transitions for the splitting processes that are allowed by the energy conservation constraint $E_{n,p_z,f}- E_{n^{\prime},p_z-k_z,f}=\Omega$ (where we set $\lambda=+1$ and $\eta =-1$) are visualized schematically in Fig.~\ref{fig:transitions}. To be specific, we concentrate on the splitting processes of up quarks as an example. The kinematics for down quark is qualitatively the same. For comparison, we show side-by-side the results for two different choices of the photon transverse momenta: a smaller value ($k_T=0.025~\mbox{GeV}$) in panel (a) and a larger value ($k_T=0.075~\mbox{GeV}$) in panel (b). Also, to get an idea about the angular dependence, the result for three different directions of the photon emission are superimposed. They are represented by color-coded arrows: $\phi=0$ (red), $\phi=\pi/6$ (green), and $\phi=\pi/2$ (blue). 

It should be noted that the allowed transitions between Landau levels can be grouped systematically into an infinite set of series $n^\prime\to n$, where $n^\prime=n+i$ and $i=1,2,3,\ldots$. For a generic angle $\phi<\pi/2$, all such series represent allowed transitions, although some of them (e.g., with large values of $i$) might be suppressed more than others. In the limiting case $\phi=\pi/2$, however, only the transitions between the adjacent Landau levels ($n+1\to n$) happen to be possible. Interestingly, this restriction does not come from the energy conservation itself. This is the consequence of the vanishing amplitude for the photon emission in the direction of the magnetic field. Mathematically, this can be understood by considering functions $\mathcal{F}_1^f$ in the limit $k_y=0$, see Eqs.~(\ref{F1-orig}), (\ref{F4-orig}), (\ref{I0-k0}), and (\ref{I2-k0}) in Appendix~\ref{all-functions}. As is easy to see,  at $k_y=0$ (which is equivalent to $\phi=\pi/2$), the corresponding functions are nonzero only when $n^\prime= n$ or $n^\prime= n\pm 1$. 

Let us now discuss how the quantization of Landau levels affects the dependence on the photon emission on the transverse momentum. As seen from Fig.~\ref{fig:transitions}(a), the quantization has a particularly profound effect on the kinematics of the allowed transitions at small values of $k_T$. The underlying reason is related to the fact that the separation between Landau levels is of the order of $\sqrt{|eB|}$ at low energies. Thus, for $k_T\ll \sqrt{|eB|}$, the transitions between quark states with low energies are impossible. In fact, as is clear from Fig.~\ref{fig:transitions}(a), the lowest lying transitions are those between quark states with large momenta $| p_z | \sim  |e_f B|/\left[k_T (1+|\sin\phi |) \right]$, where the energy separation between the adjacent Landau levels is sufficiently small. Since such transitions involve quarks with relatively large energies $E_{p_z} \gtrsim |e_f B|/\left[k_T (1+|\sin\phi |) \right]$, their contributions are  strongly suppressed by the Fermi-Dirac distribution functions in the imaginary part of the polarization tensor, see Eqs.~(\ref{Im-Pi-final}) -- (\ref{g0n}). In essence, this is the underlying mechanism that explains the suppression of the total rate when $k_T$ goes to zero, see Fig.~\ref{fig:ProdRateInt}(b). 

As the value of $k_T$ grows and becomes comparable to the Landau energy scale, $k_T\sim \sqrt{|eB|}$, the effects of quantization relax gradually. This is seen qualitatively from the representation of allowed transitions in Fig.~\ref{fig:transitions}(b), where the photon transverse momentum is three times larger (i.e., $k_T=0.075~\mbox{GeV}$). In this case, the transitions start to occur between quark states with lower energies and, as a result, the photon production rate becomes higher. Of course, eventually when $k_T\gg \sqrt{|eB|}$, the rates will start to decrease again with increasing $k_T$. Therefore, the generic behavior of the photon production rate as a function of $k_T$ is similar to that in Fig.~\ref{fig:ProdRateInt}(b). It starts growing from a very small value when $k_T\simeq 0$, reaches a maximum at certain $k_{T,max}$, and then decreases at large $k_T$.

It is interesting to note that the same quantization of Landau levels also explains the unusual ellipticity of photon emission in the region of small $k_T$, which is characterized by a negative $v_2$. Indeed, it follows from the angular dependence of the quark momenta $|p_z| \sim |e_f B|/\left[k_T (1+|\sin\phi |) \right]$ quoted earlier, which characterize the transitions between states with the lowest energies. Since the smallest value of $|p_z|$ (and, thus, the energies of quark states) is achieved at $\phi =\pi/2$, the emission rate is  largest in the corresponding direction perpendicularly to the reaction plane. On the other hand, the largest $|p_z|$ corresponds to $\phi =0$, implying that the rate is suppressed the most for the photons emitted along the reaction plane. 

The explanation of a positive $v_2$ at large $k_T$ is not as simple. One may speculate that it is analogous to a classical synchrotron radiation that is emitted predominantly in the direction perpendicular to the magnetic field. One should note, however, that there is a substantial contribution from the annihilation process, which is not a classical effect but a special feature of the relativistic plasma.   

Before concluding this section, it is instructive to discuss briefly the validity of the lowest Landau level approximation, which is often utilized when the magnetic field is strong. Formally, such an approximation can be obtained from Eq.~(\ref{Im-Pi-final}) by dropping all terms except for those with $n=n^\prime =0$. The corresponding explicit expression for $\mbox{Im} [\Pi^{\mu}_{R,\mu}]$ is presented in Appendix~\ref{app:LLL}.  As is clear, in this approximation, the quark and antiquark splitting processes (which require $n\neq n^\prime$) does not be contributing to the photon production. The only process that contributes is the quark-antiquark annihilation in the lowest Landau level. In most regimes of a hot quark-gluon plasma, with the exception of the small temperature limit ($T\ll \sqrt{|eB|}$), such an annihilation is a subdominant process, however. Therefore, one must conclude that the lowest Landau level approximation is inadequate for calculating the photon production in a hot quark-gluon plasma even if the magnetic field is very strong.

\section{Summary and Conclusions}
\label{sec:summary}

In this paper, we studied the direct photon production rate from a strongly magnetized hot quark-gluon plasma. At leading zeroth order in the coupling constant $\alpha_s$, the photons are produced by the following three types of processes: (i) the quark splitting $q\rightarrow q+\gamma $, (ii) the antiquark splitting $\bar{q} \rightarrow \bar{q}+\gamma $, and (iii) the quark-antiquark annihilation $q + \bar{q}\rightarrow \gamma$. Because of a modified energy conservation in a background magnetic field, there is no need for the gluon mediation in the underlying production mechanism. 

By analyzing the relative contribution of different processes, we found that the photon production rate is  dominated by the $1\to 2$ splitting processes in a wide range of moderately high temperatures ($T\gtrsim m_\pi$), moderately strong magnetic fields ($|eB|\gtrsim m_\pi^2$), and a range of not too large transverse momenta ($k_T \lesssim \sqrt{|eB|}$). With increasing transverse momenta, the relative contribution of the annihilation process grows and eventually becomes comparable to that of the  splitting processes. The annihilation also plays an important role in the limit of small temperature. In this connection, it is instructive to mention that the lowest Landau level approximation, which includes only the annihilation of quarks and antiquarks in the zeroth Landau level, is not reliable for calculating the photon emission from a hot plasma even if the magnetic field is strong. 

Our investigation reveals that the photon emission from a strongly magnetized hot quark-gluon plasma is characterized by a nonzero ellipticity coefficient $v_2$ that depends on the transverse momentum. Generically, $v_2$ is negative at small momenta, $k_T\lesssim \sqrt{|eB|}$, and positive at large momenta, $k_T\gtrsim \sqrt{|eB|}$. While the ellipticity coefficient $v_2$ is an overall growing function of $k_T$, it is not smooth or monotonic. This is due to the quantization of the Landau levels of quarks that produces numerous thresholds associated with the inclusion of additional quantum transitions when $k_T$ (or, equivalently, energy) increases. While the strong interaction effects in plasma are expected to smooth out the functional dependence of $v_2$, the corresponding analysis was not performed in this study. We hope to address the role of interaction effects beyond the leading order in $\alpha_s$ in the future. It is not expected, however, that they could change dramatically the overall dependence of the photon production on the transverse momentum or the angular coordinate $\phi$. 

As we found in this study, the ellipticity coefficient $v_2$ tends to saturate at large $k_T$, reaching a relatively high positive value $v_{2,max}\simeq 0.2$. Interestingly, this prediction is considerably smaller than the result in Ref.~\cite{Tuchin:2014pka}, where the use of the classical formula for synchrotron radiation gave the maximum value $v^{(class)}_{2,max}=4/7$. It appears that the effective suppression in our quantum analysis is caused by the inclusion of the annihilation processes. Nevertheless, we find that $v_2$ still approaches a moderately high value at large $k_T$. This result can have important implications for heavy-ion collisions. In particular, the magnetic field mediated processes could give a substantial contribution to the observed $v_2$ for the direct photon production. Of course, the final anisotropy could be diluted by the photon production from the standard gluon-mediated processes \cite{Kapusta:1991qp,Baier:1991em,Aurenche:1998nw,Steffen:2001pv,Arnold:2001ba,Arnold:2001ms,Ghiglieri:2013gia}, which are (mostly) isotropic in the local rest frame of the plasma. Thus, in the presence of a strong magnetic field, it is reasonable to expect that the leading order splitting and annihilation processes can be more important since they do not suffer from an extra suppression due to the Bose distribution of gluons. While the interplay of the two types of processes was not investigated in detail in this paper, it will be very interesting to address in the future.

The analysis of the $k_T$ dependence shows that the integrated photon production rate is strongly suppressed in the limit $k_T \to 0$. This is the consequence of quantum effects and the structure of quark Landau levels in a magnetic field. In essence, the physics mechanism is explained by a wide spacing of the quantized Landau levels that prohibits transitions between the low-energy states when producing photons. In fact, by analyzing the energy conservation constraint, one finds that the only allowed transitions are those involving quarks with rather high energies, i.e., $E_{p_z}\gtrsim |e_f B|/(2k_T)$. Since the number density of high-energy quarks is small, the photon production is negligible. The suppression from the Landau level quantization gets lifted gradually as the value of $k_T$ grows and the photon production starts to grow too. At certain critical point, however, it reaches a maximum and then starts to decrease with $k_T$. 

It should be emphasized that, in this study, we calculated the photon rate at the leading zeroth order in the strong coupling constant $\alpha_s$. This is expected to be a good approximation in a strongly magnetized QCD plasma. In general, however, higher-order corrections in $\alpha_s$ should exist and become increasingly important with decreasing of the magnetic field. At present, we do not know how to identify and include systematically all relevant gluon-mediated processes. One might try to include the relevant corrections by using the hard thermal loop resummations \cite{Braaten:1989mz} as done in the absence of the magnetic field \cite{Aurenche:1998nw,Steffen:2001pv}. (Note that an additional  subclass of $2 \to 3$ and $3 \to 2$ processes with a collinear enhancement become as important as the $2 \to 2$ processes \cite{Arnold:2001ba,Arnold:2001ms,Ghiglieri:2013gia}.) Unfortunately, the inclusion of strong interaction effects via the resummation of higher-order loop diagrams is far from straightforward from a technical viewpoint when the QCD plasma is magnetized. The problem stems not only from the elaborate structure of the quark propagator, but also from the intrinsic role of the magnetic field that should interfere with the thermal effects. This is suggested by the studies of the strongly magnetized QCD vacuum (i.e., $T=0$), where the resummation of the hard ``magnetic loops" becomes important \cite{Miransky:2002rp}. It leads, for example, to a nonzero (but gauge invariant) gluon mass that scales as $\sqrt{\alpha_s}\sqrt{eB}$. Similar effects are likely to survives also at nonzero temperatures and could complicate the hard thermal loop resummations. The corresponding systematic study is beyond the scope of the present paper but should be attempted in the future.

\acknowledgements
The authors thank Danning Li for providing computational resources.
The authors also thank  Bronislav Zakharov and Di-Lun Yang for useful comments about the first version of the manuscript.
The work of X.W. was supported by the start-up funding No.~4111190010 of Jiangsu University and NSFC under Grant No.~11735007.
The work of I.A.S. was supported by the U.S. National Science Foundation under Grant No.~PHY-1713950. 
L.Y. is supported by the NSFC under Grant No.~11605072 and the Seeds Funding of Jilin University. 
M.H. is supported in part by the NSFC under Grants No.~11725523, No.~11735007, No.~11261130311 (CRC 110 by DFG and NSFC), Chinese Academy of Sciences under Grant No.~XDPB09, the start-up funding from University of Chinese Academy of Sciences (UCAS), and the Fundamental Research Funds for the Central Universities.

\appendix

\section{Dirac traces and auxiliary functions}
\label{all-functions}

In this Appendix, we present the explicit expressions for the functions that appear in the calculation of the imaginary part of the (Lorentz contracted) photon polarization function. 

As is clear from the explicit structure of the quark propagators, see Eqs.~(\ref{GDn-alt}) and (\ref{Dn-quark}), there are four types of Dirac traces that appear in the calculation:
\begin{eqnarray}
g_{\mu\nu}T_{1,f}^{\mu\nu} &=&
\mbox{tr} \left[ \gamma^\mu \left(p_\parallel \gamma_\parallel + m \right)
 \left( {\cal P}_{+}L_n +{\cal P}_{-}L_{n-1} \right)
\gamma_\mu \left((p_\parallel -k_\parallel )\gamma_\parallel + m \right)
\left( {\cal P}_{+}L_{n^\prime} +{\cal P}_{-}L_{n^\prime-1} \right)\right]  \nonumber\\
&=&4\left[m^2- p_\parallel (p_\parallel -k_\parallel ) \right]
\left(L_{n-1}L_{n^\prime}+L_{n}L_{n^\prime-1}\right)
+4 m^2\left(L_{n}L_{n^\prime}+L_{n-1}L_{n^\prime-1}\right)
\label{trt1-cross},\\
g_{\mu\nu}T_{2,f}^{\mu\nu} &=& \frac{i}{l_f^2}
\mbox{tr} \left[ \gamma^\mu \left(p_\parallel \gamma_\parallel + m \right)\left( {\cal P}_{+}L_n +{\cal P}_{-}L_{n-1} \right)
\gamma_\mu  (\mathbf{r}_{\perp}\cdot\bm{\gamma}_{\perp}) L_{n^\prime-1}^{1} \right] = 0
\label{trt2-cross},\\
g_{\mu\nu}T_{3,f}^{\mu\nu} &=& -\frac{i}{l_f^2}
\mbox{tr} \left[ \gamma^\mu (\mathbf{r}_{\perp}\cdot\bm{\gamma}_{\perp}) L_{n-1}^{1}     
\gamma_\mu  \left((p_\parallel -k_\parallel )\gamma_\parallel + m \right)
\left( {\cal P}_{+}L_{n^\prime} +{\cal P}_{-}L_{n^\prime-1} \right)\right] = 0
\label{trt3-cross},\\
g_{\mu\nu}T_{4,f}^{\mu\nu} &=&\frac{1}{l_f^4}
\mbox{tr} \left[ \gamma^\mu (\mathbf{r}_{\perp}\cdot\bm{\gamma}_{\perp}) L_{n-1}^{1}    
\gamma_\mu  (\mathbf{r}_{\perp}\cdot\bm{\gamma}_{\perp}) L_{n^\prime-1}^{1} \right]
= \frac{8 }{l_f^4}  \mathbf{r}_\perp^2  L_{n-1}^{1}  L_{n^\prime-1}^{1} ,
\label{trt4-cross}
\end{eqnarray}
where, for brevity of notation, the argument $\xi = \mathbf{k}_{\perp}^2l_f^{2}/2$ of the Laguerre polynomials is suppressed. 

After  the integration over the transverse spatial coordinates, these produce the following functions:
\begin{eqnarray}
\mathcal{F}_1^f&=&  g_{\mu\nu}  I_{1,f}^{\mu\nu} 
=\int   d^2 \mathbf{r}_\perp e^{-i \mathbf{r}_\perp\cdot \mathbf{k}_\perp} e^{-\mathbf{r}_\perp^2/(2l_f^2)} g_{\mu\nu}T_{1,f}^{\mu\nu} 
\nonumber\\
&=&8\pi l_f^2 \left[m^2- p_\parallel(p_\parallel -k_\parallel) \right]
\left( \mathcal{I}_{0,f}^{n-1,n^{\prime}}(\mathbf{k}_\perp) +\mathcal{I}_{0,f}^{n,n^{\prime}-1}(\mathbf{k}_\perp)  \right)
+8\pi l_f^2m^2\left( \mathcal{I}_{0,f}^{n,n^{\prime}}(\mathbf{k}_\perp) +\mathcal{I}_{0,f}^{n-1,n^{\prime}-1}(\mathbf{k}_\perp)  \right)
\label{F1-orig},\\
\mathcal{F}_4^f&=&g_{\mu\nu}  I_{4,f}^{\mu\nu} 
= \int   d^2 \mathbf{r}_\perp e^{-i \mathbf{r}_\perp\cdot \mathbf{k}_\perp} e^{-\mathbf{r}_\perp^2/(2l_f^2)} g_{\mu\nu}T_{4,f}^{\mu\nu} = 16 \pi \, \mathcal{I}_{2,f}^{n-1,n^{\prime}-1}(\mathbf{k}_\perp).
\label{F4-orig}
\end{eqnarray}
(Note that $\mathcal{F}_2^f=\mathcal{F}_3^f=0$.) Since function $\mathcal{F}_1^f$ depends explicitly on the zeroth component of the fermion four-momentum, it has to be treated with care when the Matsubara summation is performed. In effect, the Matsubara sum produces the result which is equivalent to the following replacement: 
\begin{equation}
p_\parallel  (p_\parallel -k_\parallel) \to  \lambda E_{n,p_z,f}E_{n^{\prime},p_z-k_z,f} -p_z(p_z-k_z) .
\label{reslt-1}
\end{equation}
Furthermore, when the fermion energies satisfy the energy conservation condition $E_{n,p_z,f}-\lambda E_{n^{\prime},p_z-k_z,f}+\eta \Omega =0$, one finds that
\be  
\lambda E_{n,p_z,f}E_{n^{\prime},p_z-k_z,f} -p_z(p_z-k_z)  = m^2+(n+n^\prime)|e_f B| +\frac{1}{2}\left(k_z^2-\Omega^2\right).
\label{reslt-1}
\end{equation}
Thus, in the calculation of the imaginary part of the polarization function, it is convenient to use the following expression for function $\mathcal{F}_1^f$: 
\begin{equation}
\mathcal{F}_1^f =
8\pi \left[\frac{\Omega^2-k_z^2}{2|e_f B| } -(n+n^\prime)\right]
\left( \mathcal{I}_{0,f}^{n-1,n^{\prime}}(\mathbf{k}_\perp) +\mathcal{I}_{0,f}^{n,n^{\prime}-1}(\mathbf{k}_\perp)  \right)
+8\pi l_f^2 m^2\left( \mathcal{I}_{0,f}^{n,n^{\prime}}(\mathbf{k}_\perp) +\mathcal{I}_{0,f}^{n-1,n^{\prime}-1}(\mathbf{k}_\perp)  \right),
\end{equation}
which is equivalent to Eq.~(\ref{F1-orig}) provided $E_{n,p_z,f}-\lambda E_{n^{\prime},p_z-k_z,f}+\eta \Omega =0$.

The $\mathcal{F}_i^f$ functions are expressed in terms of the following two functions:
\begin{eqnarray}
\mathcal{I}_{0,f}^{n,n^{\prime}}(\mathbf{k}_\perp)&=& \frac{(n^\prime)!}{n!} e^{-\xi}  \xi^{n-n^\prime}
\left(L_{n^\prime}^{n-n^\prime}\left(\xi\right)\right)^2 
= \frac{n!}{(n^\prime)!}e^{-\xi} \xi^{n^\prime-n}
\left(L_{n}^{n^\prime-n}\left(\xi\right)\right)^2 ,
\label{I0f-LL-form2} \\
\mathcal{I}_{2,f}^{n,n^{\prime}}(\mathbf{k}_\perp)&=& 2 \frac{(n^\prime+1)!}{n!}e^{-\xi}  \xi^{n-n^\prime} 
L_{n^\prime}^{n-n^\prime}\left(\xi\right)
L_{n^\prime+1}^{n-n^\prime}\left(\xi\right) 
= 2 e^{-\xi} \frac{(n+1)!}{(n^\prime)!}\xi^{n^\prime-n}  L_{n}^{n^\prime-n}\left(\xi\right)L_{n+1}^{n^\prime-n}\left(\xi\right)  .
\label{I2f-LL-form2} 
\end{eqnarray}
Note that, for each function, there are two formally different but mathematically equivalent representations. In
numerical calculations, however, the evaluation errors could be minimized by using the first form when 
$n> n^\prime$ and the second when $n< n^\prime$. 

By using the properties of the Laguerre polynomials \cite{Gradshtein}, the following asymptotic behavior 
of $\mathcal{I}_{0,f}^{n,n^{\prime}}$ and $\mathcal{I}_{2,f}^{n,n^{\prime}}$ can be derived:
\begin{eqnarray}
\mathcal{I}_{0,f}^{n,n^{\prime}}(\mathbf{k}_\perp) &\simeq&  \delta_{n,n^{\prime}}
- \frac{1}{2}\left[(2n+1)\delta_{n,n^{\prime}} -(n+1)\delta_{n,n^{\prime}-1}
- (n^{\prime}+1)\delta_{n-1,n^{\prime}}\right]\left(\mathbf{k}_{\perp}l_f\right)^2
+O\left[\left(\mathbf{k}_{\perp}l_f\right)^4\right],
\label{I0-k0}
\\
\mathcal{I}_{2,f}^{n,n^{\prime}}(\mathbf{k}_\perp)  &  \simeq & 2 (n+1)  \delta_{n,n^{\prime}}
 - (n+1) (n^{\prime}+1)\left(2\delta_{n,n^{\prime}}- \delta_{n,n^{\prime}-1}-  \delta_{n-1,n^{\prime}} \right)
\left(\mathbf{k}_{\perp}l_f\right)^2 +O\left[\left(\mathbf{k}_{\perp}l_f\right)^4\right],
\label{I2-k0}
\end{eqnarray}
in the limit of  small $|\mathbf{k}_{\perp}| l_f$.

\section{Lowest Landau level approximation}
\label{app:LLL}

In the lowest Landau level approximation, the explicit result for the Lorentz-contracted imaginary part of the polarization tensor follows from Eq.~(\ref{Im-Pi-final}) by omitting all terms with $n$ and $n^{\prime}$ larger than $0$. The corresponding result reads
\begin{eqnarray}
\mbox{Im} \left[\Pi^{\mu}_{\mu}\right] &=&
\frac{4 N_c  m^2 \Theta\left(k_y^2 -4m^2\right)}{k_y^2 R_m} 
\sum_{f=u, d} \frac{\alpha_f}{l_f^2}  e^{-k_y^2 l_f^{2}/2} 
\left[ n_F\left(\frac{\Omega-k_z R_m}{2}\right)+n_F\left(\frac{\Omega+k_z R_m}{2}\right)-1 \right]  ,
\label{ImPi-LLL}
\end{eqnarray}
where $R_m = \sqrt{1-4m^2/k_y^2 }$ and $\Omega = \sqrt{k_y^2+k_z^2}$. 

As is easy to check, the result in Eq.~(\ref{ImPi-LLL}) is consistent with the spectral function obtained in
the lowest Landau level approximation in Ref.~\cite{Bandyopadhyay:2016fyd}. As emphasized in the main text, however, this 
approximation is not very reliable for calculating the photon production rate even in the case of very strong magnetic fields.

\end{document}